\documentclass[onecolumn,useAMS,usenatbib]{mn2e}

\usepackage{color}
\usepackage{amsmath}
\usepackage{graphicx}
\usepackage{amssymb}
\usepackage{psfrag}

\newcommand{\beq}{\begin{equation}}
\newcommand{\eeq}{\end{equation}}
\newcommand{\bay}{\begin{array}}
\newcommand{\eay}{\end{array}}
\newcommand{\beqy}{\begin{eqnarray}}
\newcommand{\eeqy}{\end{eqnarray}}

\newcommand{\rmd}{\mathrm{d}}
\newcommand{\rmp}{\mathrm{p}}
\renewcommand{\rmn}{\mathrm{n}}
\newcommand{\rmx}{\mathrm{x}}
\newcommand{\brac}[1]{\left({#1}\right)}

\newcommand{\pd}[2]{\frac{\partial{#1}}{\partial{#2}}}
\newcommand{\td}[2]{\frac{\rmd{#1}}{\rmd{#2}}}
\newcommand{\curl}{\nabla\times}
\renewcommand{\div}{\nabla\cdot}
\renewcommand{\lor}{\boldsymbol{\mathcal{L}}}
\newcommand{\bj}{{\bf j}}
\newcommand{\bB}{{\bf B}}
\newcommand{\pom}{\varpi}
\newcommand{\be}{{\bf e}}
\newcommand{\br}{{\bf r}}
\newcommand{\ve}{\varepsilon}
\newcommand{\eps}{\epsilon}
\newcommand{\Aph}{A_\phi}
\newcommand{\mup}{\tilde{\mu}_\rmp}
\newcommand{\mun}{\tilde{\mu}_\rmn}
\newcommand{\clE}{\mathcal{E}}
\newcommand{\clM}{\mathcal{M}}
\newcommand{\Bav}{\bar{B}}
\newcommand{\scl}{\boldsymbol{\mathfrak{F}}_{\mathrm{mag}}}
\newcommand{\uj}{\boldsymbol{\hat{\j}}}
\newcommand{\fmag}{{\bf F}_{\mathrm{mag}}}
\newcommand{\rhom}{\rho_{\mathrm{max}}}
\newcommand{\emag}{\clE_{\mathrm{mag}}}
\newcommand{\etor}{\clE_{\mathrm{tor}}}
\newcommand{\Btor}{B_{\mathrm{tor}}}
\newcommand{\Bpol}{B_{\mathrm{pol}}}

\newcommand{\skl}[1]{{\color{black}{#1}}}

\title[Stratified and superconducting NS equilibria]{Magnetic neutron
  star equilibria with stratification and type-II superconductivity}
\author[S. K. Lander, N. Andersson and K. Glampedakis]
       {S. K. Lander${}^{1,2}$\thanks{skl@soton.ac.uk},
         N. Andersson${}^1$ and K. Glampedakis${}^3$\\ \\
${}^1$ Mathematical Sciences, University of Southampton, Southampton
       SO17 1BJ, U.K.\\
${}^2$ Max-Planck-Institut f\"ur Gravitationsphysik,
Albert-Einstein-Institut, Potsdam-Golm 14476, Germany\\
${}^3$ Theoretical Astrophysics, University of T\"{u}bingen, Auf der
Morgenstelle 10, T\"{u}bingen 72076, Germany}

\begin{document}


\pagerange{\pageref{firstpage}--\pageref{lastpage}} \pubyear{0000}

\maketitle
\vspace{-0cm}

\label{firstpage}

\begin{abstract}
We construct two-fluid equilibrium configurations for
neutron stars with magnetic fields, using a self-consistent
and nonlinear numerical approach. The two-fluid approach
\skl{--- likely to be valid for large regions of all but the
  youngest NSs ---}
provides us with a straightforward way to introduce stratification and
allows for more realistic models than the ubiquitous barotropic
assumption. In all our models the neutrons are modelled as a superfluid,
whilst for the protons we consider two cases: one where they are a normal
fluid and another where they form a type-II superconductor. We
consider a variety of field configurations in the normal-proton case
and purely toroidal fields in the superconducting case. We find that
stratification allows for a stronger toroidal component in mixed-field
configurations, though the poloidal component remains the largest
\skl{in all our models}.
We provide quantitative results for magnetic ellipticities
of NSs, both in the normal- and superconducting-proton cases.
\end{abstract}


\section{Introduction}

It has long been acknowledged that neutron stars (NSs) have strong
magnetic fields, which may play a variety of roles in the physics of
these objects. Magnetic fields can change the stability of a
NS, affect its evolution and cooling properties, induce precession and
provide the restoring force for a class of oscillation modes
\citep{harding_lai,mereghetti}. They may also produce crust-core
coupling and affect the post-glitch spin behaviour of NSs
\citep{easson,dib_kg}. All of these effects are likely to be sensitive
to the details of the 
magnetic-field configuration in the stellar interior, whereas we are
only able to observe the exterior, dominantly dipolar, field. For this
reason considerable effort has gone into modelling the interior
magnetic fields of neutron stars; we give an overview of this in
section 2.

Despite considerable progress, the vast majority of studies
on magnetised neutron stars treat them as barotropic single-fluid
bodies. Whilst this is a sensible way to begin 
producing simplified models, more realistic studies need to confront
some of the detailed interior physics of NSs. For example, over forty
years since the first predictions that neutron
star matter would contain superfluid neutrons and superconducting
protons (see, e.g., \citet{baym_pp}), there have been no attempts to
construct multifluid 
magnetic equilibria along these lines. The issue is highly topical,
too --- recent observations of the cooling of the young NS in Cassiopeia A
are well explained by a model with superfluid neutrons and
superconducting protons \citep{casA_sht,casA_page}.
In the case of no magnetic field there has been more progress,
including studies of the equilibria and oscillation modes of
superfluid NSs (e.g. \citet{prix_r,prix_nc,yosh_eri,passa_sf}). The
closest related work on magnetised stars is probably the study of
\emph{single-fluid} barotropic NSs with superconductivity, by
\citet{akgun_wass}.

A simpler problem than a NS with superfluidity/superconductivity is a
stratified (i.e. non-barotropic), magnetised star composed of normal
fluid. Stratification is expected to exist in NSs \citep{reis_gold},
where it can 
allow for a wider range of equilibrium configurations as well as
providing a stabilising effect on the magnetic field
\citep{reis_strat}. Again, the limitations of barotropic stellar
models have been known for decades \citep{mestel_nonbar}, but there
have been few attempts to construct stratified magnetised
equilibria. Most of the work in this direction has been that of
Braithwaite and collaborators, who perform nonlinear
magnetohydrodynamics (MHD) evolutions of
a stratified star and show that the system appears to relax to an
equilibrium over time \citep{braith_nord,braithtorpol}. \skl{In addition,
\citet{mastrano} very recently studied single-fluid non-barotropic
models of NSs, albeit using pre-specified magnetic field configurations.}

In this paper we attempt to plug some of the gaps in the
literature, presenting the first results for magnetised NS 
equilibria modelled as a two-fluid system. We treat the neutrons as
being a superfluid, which mainly interacts with the proton fluid through the
star's gravitational potential. By choosing different degrees of
compressibility for the two fluid species, we can introduce
stratification. We begin by considering the case where protons are a
magnetised normal fluid and find 
equilibria with purely poloidal, purely toroidal and mixed-field
configurations. After this, we treat the `full' problem where the
neutron star is composed of superfluid neutrons and type-II
superconducting protons for the first time, specialising to the case
of a purely toroidal magnetic field. This paper takes a nonlinear
approach to the problem, but we have also worked on a separate study
using perturbation theory; the results of this are presented in a
parallel paper, \citet{GAL}.

This paper is laid out as follows. In section 2 we present the
equilibrium equations for our stellar model, independent of the
details of the magnetic field. We then specialise to the case where
the protons are a normal fluid governed by standard MHD. In section 3
we consider the case where the protons are instead a type-II
superconductor, presenting simplified equations which nonetheless are
likely to be accurate enough for an initial study of equilibria. In the case of
purely toroidal fields we derive equations in a form suitable for
numerical integration. In section 4 we give details of our numerical
procedure, whilst in section 5 we show how to convert code output (in
terms of dimensionless code variables) into useful physical
quantities. Section 6 contains our results for two-fluid
equilibria, exploring the effect of stratification and type-II
superconductivity and section 7 is a discussion.

\section{Two-fluid equilibrium equations and normal MHD}

\subsection{Equilibrium equations for two-fluid magnetic stars}

We begin by modelling a neutron star as axisymmetric, with the
magnetic symmetry axis being the same as the rotation axis. For this
system it is natural to use cylindrical polar coordinates
$(\pom,\phi,z)$, where the $z$-axis is aligned with the symmetry axis
of the star. Although relativistic effects will certainly play a role
in neutron star physics, we will neglect these for this study, and
work in Newtonian gravity. We also ignore the elastic crust
and any non-hadronic matter that may form an inner core, and assume
that the neutron star is a multifluid system, with a neutral fluid of
neutrons and a charged fluid of protons and electrons.

\skl{Our multifluid model should be applicable for the bulk of a
  neutron star's interior during much, but not all, of its life. A
  neutron star is born hot, with its charged and neutral components
  strongly coupled through various dissipative mechanisms; at this
  stage it behaves as a  \emph{single}, 
  non-barotropic, fluid. When the star cools below a critical
  temperature $T_c$ however, the core neutrons begin to condense into
  a superfluid 
  state\footnote{\skl{Neutron superfluidity in the inner crust is likely to
      occur at a higher temperature --- above $10^9$ K --- due to its
      singlet pairing type (as opposed to the triplet type in the core),
      making this part of the star a multifluid system too.}}. The
  recent observations of the cooling of the Cassiopeia A NS indicate
  that $T_c\sim (5-9)\times 10^8$ K; a NS core should start to drop
  below this temperature $\sim 100$ years after birth
  \citep{casA_page,casA_sht}. For this 
  reason, even young pulsars are likely to have substantial multifluid
  regions --- i.e., regions where the neutrons are superfluid. Models
  of magnetar temperature profiles \citep{kaminker_magprof} indicate
  they are cool enough to have 
  multifluid regions too. For this preliminary study into stratified
  NSs we model the \emph{entire} star as multifluid; this is obviously
  a simplification, and more detailed
  modelling should account for the layered structure (including
  single-fluid regions) expected in real NSs.}

Now, since electrons have a much smaller mass than either of the other
components, we take the standard approach of neglecting their
inertia. Our final model is
then a two-fluid system, with a proton fluid (whose variables are
denoted with a subscript roman index p) and a neutron fluid (with
index n). Neutrons and protons are assumed to be of equal mass
$m_\rmn=m_\rmp\equiv m$. Next, we denote the particle chemical
potentials by $\mu_\rmn$ and $\mu_\rmp$, then define $\mun\equiv\mu_\rmn/m$ and
$\mup\equiv(\mu_\rmp+\mu_\mathrm{e})/m$; note that we use $\mup$ to
denote the chemical potential of the whole charged component, as it
also contains a contribution from the electrons $\mu_\mathrm{e}$.

In normal MHD, only the protons are coupled to the magnetic field, but
for a superfluid/superconducting system there is 
generally a magnetic force acting on the neutrons as well
\citep{mendell,GAS2011}. In this work however, we only consider the case
without entrainment, for which the neutron superfluid is decoupled
from the magnetic field in the superconducting case too. This issue is
discussed in more detail in \citet{GAL}. 
Given these assumptions, we may now state the separate 
Euler equations which govern the two fluids:
\beq \label{n_Euler}
\nabla\brac{\mun+\Phi-\frac{\pom^2\Omega_\rmn^2}{2}}=0,
\eeq
\beq \label{p_Euler}
\nabla\brac{\mup+\Phi-\frac{\pom^2\Omega_\rmp^2}{2}}=\frac{\fmag}{\rho_\rmp},
\eeq
where $\fmag$ denotes the magnetic force; the form of
this force will be the only difference between the normal-MHD and
superconducting cases to our order of working. As usual, $\Phi$ denotes the
gravitational potential, $\rho$ is density and $\Omega$ denotes
rotation rate (with indices for individual fluid species). We will
only consider the case where there is no 
entrainment and the two fluids corotate,
i.e. $\Omega_\rmn=\Omega_\rmp\equiv\Omega$. Since we are only studying
stationary configurations, 
the equations do not contain mutual friction terms.

As in the single-fluid case we have Poisson's equation for the
gravitational potential, from which we see that the behaviour of the
two fluids is linked despite the neutrons being a superfluid:
\beq
\nabla^2\Phi = 4\pi G\rho = 4\pi G(\rho_\rmn+\rho_\rmp).
\eeq
Let us multiply equation \eqref{n_Euler} by $\rho_\rmn$, equation
\eqref{p_Euler} by $\rho_\rmp$, and add them:
\beq
\rho_n\nabla\mun+\rho_\rmp\nabla\mup
 +\rho\nabla\Phi-\rho\nabla\brac{\frac{\pom^2\Omega^2}{2}}=\fmag.
\eeq
Now since
$\rho_n\nabla\mun+\rho_p\nabla\mup = \nabla P$ (the gradient of the
total fluid pressure), this allows us to recover the familiar
`single-fluid' Euler equation:
\beq
\frac{\nabla P}{\rho}+\nabla\Phi-\nabla\brac{\frac{\pom^2\Omega^2}{2}}
  = \frac{\fmag}{\rho},
\eeq
except that in this two-fluid case one will typically have
stratification, and hence can no longer replace the pressure term with an
enthalpy gradient. Instead, it is more useful to work with the
proton-fluid Euler and a `difference-Euler', given 
by \eqref{p_Euler} minus \eqref{n_Euler}:
\beq \label{d_Euler}
\nabla\brac{\mup-\mun} = \frac{\fmag}{\rho_\rmp}.
\eeq
Taking the curl of this equation, we see that (as in the single-fluid
case) there exists a scalar $M$ such that
\beq \label{M_defn}
\frac{\fmag}{\rho_\rmp}=\nabla M.
\eeq
The magnetic force --- whether the normal MHD `Lorentz force' or the
flux-tube tension force of a type-II superconductor (see section 3)
--- therefore depends on a scalar function $M$ and the 
proton-fluid density $\rho_\rmp$, as opposed to the single-fluid case
where the dependence is on the total mass density: $\fmag/\rho=\nabla
M$. In deriving the equations that govern the behaviour of the
magnetic field in the two-fluid case, the only differences from the
single-fluid case will be in this density factor; the final magnetic
equations will be the same but with $\rho$ replaced by $\rho_\rmp$.

For our numerical scheme we need to work with the equilibrium
equations in integral form; in this section we discuss those equations
which hold for all magnetic-field configurations. Firstly, we always
have the same form of Poisson's equation for the gravitational potential:
\beq \label{int_Poisson}
\Phi(\br) 
  = -G \int \frac{\rho_\rmn(\br')+\rho_\rmp(\br')}{|\br-\br'|}\ \rmd\br'.
\eeq
We also have first-integral forms of the proton-fluid Euler
equation \eqref{p_Euler}:
\beq \label{pEuler}
\mup+\Phi-\frac{\pom^2\Omega^2}{2} = M+C_\rmp
\eeq
and the difference-Euler \eqref{d_Euler}:
\beq \label{dEuler}
\mup-\mun = M + C_d,
\eeq
where $C_\rmp$ and $C_d$ are the integration constants for the proton
and difference-Euler equations, respectively.
Finally, for the equation of state we choose an energy
functional $\clE$ given by
\beq \label{eos_functional}
\clE = k_\rmn\rho_\rmn^{\gamma_\rmn}+k_\rmp\rho_\rmp^{\gamma_\rmp},
\eeq
a two-fluid analogue of a polytropic model
\citep{prix_ca,passa_sf}. This could also be written 
in terms of the polytropic indices $N_\rmx=1/(\gamma_\rmx-1)$,
$\rmx=\{\rmn,\rmp\}$. Note that \eqref{eos_functional} can be
generalised to include cross-terms, corresponding to symmetry energy
and entrainment. The chemical potentials are then defined from the
energy functional by 
\beq \label{mux_defn}
\tilde\mu_\rmx\equiv\left.\pd{\clE}{\rho_\rmx}\right|_{\rho_{\mathrm{y}}},
\eeq
where the index x represents one particle species (neutrons or
protons) and the y index represents the other. The other equations of
the system depend on the details of the magnetic-field configuration;
we discuss these next.

\subsection{Normal MHD}

The simplest case is when the charged component of the stellar matter
is governed by normal MHD. The magnetic force $\fmag$ is then the
familiar Lorentz force, given by
\beq \label{lor_force}
\lor=\bj\times\bB=\frac{1}{4\pi}(\curl\bB)\times\bB,
\eeq
where $\bB$ is the magnetic field and $\bj$ the current.
Most studies of NS equilibria have used the normal-MHD equations and
have additionally adopted a single-fluid model, where the charged
component is the entire fluid --- this ansatz has been used by
innumerable authors. Recent examples include
\citet{tomi_eri,kiuchi,haskell,1f_eqm} and \citet{ciolfi}, but there
are decades of earlier work on single-fluid models --- see, e.g.,
\citet{ferraro} and \citet{roxburgh}.

An obvious way to progress is to work with a two-fluid model, but
still in normal MHD: the protons are a normal fluid and the neutrons
superfluid. This is not just a stepping stone towards the
superfluid/superconducting model of \citet{baym_pp} --- it already allows
us to study the role of stratification in magnetised
stars. Furthermore, the interior magnetic field strength in magnetars
may well exceed the second critical field $H_{c2}\sim 10^{16}$ G at
which superconductivity is destroyed; in this case magnetar matter
could indeed be predominantly normal protons and superfluid neutrons
\citep{GAL}.

Since only the proton fluid is magnetised, the derivations for all
equations with magnetic terms proceed in the same manner as the
single-fluid case, with factors of $\rho$ replaced by $\rho_\rmp$;
otherwise they are identical. For this reason we do not give the
details here, but the essentially identical single-fluid derivation
may be found in various papers (see, e.g., \citet{1f_eqm}). The 
non-magnetic equations of the system are those discussed in the
previous subsection.

In the mixed-field (or purely poloidal field) case we have a Poisson
equation for the $\phi$-component of the magnetic vector potential
too, \skl{allowing for the magnetic field to extend outside the star}:
\beq \label{int_Aph}
\Aph(\br)\sin\phi 
  = \frac{1}{4\pi} 
      \int\ \frac{ \frac{1}{\tilde\pom}f(\tilde u)f'(\tilde u)
                   + 4\pi\tilde\pom M'(\tilde u)\rho_\rmp(\tilde\br)
                 }{|\br-\tilde\br|}
           \ \sin\tilde\phi\ \rmd\tilde\br,
\eeq
where $u$ is a streamfunction; see the next section. In this
equation $f$ and $M$ are functions of $u$, tildes denote dummy
variables and primes denote 
derivatives. In addition, $u$ and $\Aph$ are related through
$u=\pom\Aph$, so the integral in equation \eqref{int_Aph} may be
rewritten in terms of $\Aph$.  As mentioned earlier, this integral
equation for $\Aph$ is identical to the single-fluid version
except with a $\rho$ term in the integrand replaced by $\rho_\rmp$.

\skl{The magnetic functions $M(u)$ and $f(u)$ appear to be arbitrary, but
in practice we have found only one acceptable functional form for
each; this was also the case in our earlier study of single-fluid
models \citep{1f_eqm}. In the mixed-field case, we use the following
form of $M$:}
\beq
M_{mix} = \kappa u = \kappa\pom\Aph
\eeq
where $\kappa$ is a constant governing the strength of the
Lorentz force. \skl{We find that choosing $M(u)$ as a lower or higher power
of $u$ results in the code, through equation \eqref{int_Aph}, iterating 
to an unmagnetised equilibrium solution --- and hence only the above
choice of $M_{mix}$ is acceptable in our work.}

The other magnetic function $f$ in the integral equation for
$\Aph$ relates to the toroidal component of the field and is defined
to be non-zero only within the largest interior closed field line, so there is
no exterior toroidal field \skl{(and hence no exterior current)}:
\beq \label{f_defn}
f(u)=a(u-u_{\max})^{1.1} H(u-u_{\max}),
\eeq
where $u_{max}$ is the maximum value attained by $u$ within the star
and $H$ is the Heaviside step function. The strength of the toroidal
component is adjusted by varying the constant $a$. Other choices of
the exponent than $1.1$ are possible; we choose this value as it gives
a relatively strong toroidal-field component. Choosing a value of
unity or less for the exponent gives undesirable behaviour in $f'(u)$,
however \citep{1f_eqm}. \skl{Finally, whilst adjusting the exponent in
\eqref{f_defn} allows for configurations with very modest differences,
we do not know of any genuinely different functional form of $f$
which is still dependent on $u$ and restricts the toroidal component to the
stellar interior.}

In the toroidal-field case there is no separate integral equation for the
magnetic field. $M$ has a different functional form in the
pure-toroidal field case $M_{tor}$ from that in the
mixed/pure-poloidal field case $M_{mix}$. As with other quantities,
the two-fluid form of $M_{tor}$ may be found from the single fluid
version by replacing $\rho$ with $\rho_\rmp$:
\beq
M_{tor}
  = -\frac{1}{4\pi}\int_0^{\rho_\rmp\pom^2}
                      \frac{h(\zeta)}{\zeta}\td{h}{\zeta}\ \rmd\zeta,
\eeq
where $h$ is a function of $\rho_\rmp\pom^2$, related to the magnetic
field by
\beq
B_\phi=\frac{h(\rho_\rmp\pom^2)}{\pom}.
\eeq
We choose $h(\rho_\rmp\pom^2)=\lambda\rho_\rmp\pom^2$, where $\lambda$
is a constant governing the strength of the toroidal field. \skl{Once
  again, all other functional forms appear to result in either
  divergent magnetic quantities or our numerical scheme iterating to
  an unmagnetised solution \citep{1f_eqm}.}

Note that the difference-Euler \eqref{d_Euler} for an unmagnetised
star shows that the two fluids are in chemical equilibrium. In the
magnetic case there is an extra $\fmag$ term --- this could be
interpreted as a statement that the star is out of chemical
equilibrium \citep{GAL}. \skl{Alternatively, one could adjust the notion of
what is meant by `chemical equilibrium'. Since the energy functional
$\clE$ gains an extra contribution from the presence of a magnetic field
\citep{GAS2011}, the chemical potentials --- defined through
\eqref{mux_defn} --- also change accordingly. One could then
see equation \eqref{d_Euler} as quantifying the difference
between magnetised and unmagnetised chemical equilibria.}

\section{Magnetic forces in type-II superconductivity}

In a typical NS the protons are likely to form a type-II superconductor,
where the field penetrates the star through thin fluxtubes
\citep{baym_pp}. This results in a magnetic force dependent on the averaged
magnetic field $\bB$ but also on the first critical field
$H_{c1}$. The result is a fluxtube tension force 
\citep{easson_peth,mendell,GAS2011}, quite 
different from the Lorentz force \eqref{lor_force}.

The aim of this section is not to give a complete description of the
equations governing a star with type-II superconducting protons, but
instead to explore how many of the steps from the normal-fluid
Grad-Shafranov derivation (see, e.g., \citet{1f_eqm}) also hold in this
case. We derive a general `interim' expression 
for the magnetic field, analogous to the normal-MHD case, at which
point one has to specialise to either purely toroidal fields or mixed
fields (with purely poloidal fields as a particular limiting
case). A key result for mixed fields from the normal-MHD case no
longer holds in the superconducting case, and we defer a detailed
study of this to a later paper. Instead we concentrate on purely
toroidal magnetic fields, where the derivation is more straightforward.

\subsection{General form}

In \citet{GAS2011}, a full derivation of the equations governing
a type-II superconducting neutron star is presented. For
our purposes, however, only a reduced and simplified set are
needed. We consider 
the case where there is no magnetic force on the neutrons, and neglect
small terms like the `London field'. Given this, our equations reduce
to those of section 2.1, 
by setting $\fmag$ to be the superconducting magnetic force
$\scl$. Before discussing this force, we have one useful result from
normal MHD which still holds in this case. Using the fact that
$\div\bB=0$ together with the assumption 
of axisymmetry allows us to write $\bB$ in terms of a streamfunction
$u$ \citep{1f_eqm}:
\beq
\bB \equiv \bB_{pol}+\bB_{tor} = \frac{1}{\pom}\nabla u\times\be_\phi +
          B_\phi\be_\phi.
\eeq
As in normal MHD then, we have $\bB\cdot\nabla u=0$.
At this point we turn to equation (153) from \citet{GAS2011} for the
explicit form of $\scl$:
\beq \label{orig_GAS_force}
\scl = \frac{1}{4\pi}
                        \Big[
                          (\bB\cdot\nabla)(H_{c1}\hat\bB)-B\nabla H_{c1}
                        \Big]
                   -\frac{\rho_\rmp}{4\pi}\nabla\brac{B\pd{H_{c1}}{\rho_\rmp}},
\eeq
\skl{where $\hat\bB$ is a magnetic unit vector and $B$ is the
(position-dependent) magnitude of the field, so that $\bB=B\hat\bB$.
From the calculation in appendix A2 of \citet{GAS2011}, the first critical
field $H_{c1}=h_c\rho_\rmp/\ve_\star$ to a good approximation, where
$\ve_\star$ is entrainment and $h_c$ a constant parameter. Since 
we are assuming that there is no neutron-fluid magnetic force, and
hence no entrainment, we have $\ve_\star=1$.} The magnetic force then becomes
\beq \label{orig_fmag}
\scl = \frac{h_c}{4\pi}
               \left[
                 (\bB\cdot\nabla)(\rho_\rmp\hat{\bB})
                 -\nabla(B\rho_\rmp)
               \right].
\eeq
Now by using a vector identity to rewrite the first term on the RHS and
rearranging the result we arrive at:
\beq \label{brack_term}
-\frac{4\pi}{h_c}\scl 
   = \rho_\rmp\nabla B 
      + \bB\times\brac{ \rho_\rmp\curl\hat\bB
                                    + \nabla\rho_\rmp\times\hat\bB }.
\eeq
Next we want to rewrite the bracketed term on the RHS of equation
\eqref{brack_term}. Let us start by 
defining a `unit current' $\uj\equiv\curl\hat\bB$; as in the
single-fluid case it may be shown that
\beq
\uj = \frac{1}{\pom}\nabla(\pom\hat{B}_\phi)\times\be_\phi
            + \hat{\j}_\phi\be_\phi.
\eeq
We then decompose both of the bracketed terms from \eqref{brack_term}
into poloidal and toroidal components. The resultant expression may
then be rearranged to show that
\beq
\rho_\rmp\curl\hat\bB + \nabla\rho_\rmp\times\hat\bB
 = \frac{1}{\pom}\nabla(\rho_\rmp\pom\hat{B}_\phi)\times\be_\phi
 + \brac{\rho_\rmp\hat\j_\phi
               - \frac{\nabla\rho_\rmp\cdot\nabla u}{\pom B} }\be_\phi
\eeq
The magnetic force becomes
\beq \label{2and3RHS}
-\frac{4\pi}{h_c}\scl 
   = \rho_\rmp\nabla B
        +\frac{1}{\pom}\bB
             \times\bigg(\nabla(\rho_\rmp\pom\hat{B}_\phi)\times\be_\phi\bigg)
        +\brac{\rho_\rmp\hat\j_\phi
               - \frac{\nabla\rho_\rmp\cdot\nabla u}{\pom B} }\bB\times\be_\phi.
\eeq
Again, it is useful to rearrange some terms in this expression. The
second term on the RHS of \eqref{2and3RHS} may be rewritten as:
\beq
\frac{1}{\pom}\bB\times
 \brac{\nabla(\rho_\rmp\pom\hat{B}_\phi)\times\be_\phi}
  = \frac{1}{\pom^2}\nabla u\times
              \nabla(\rho_\rmp\pom\hat{B}_\phi)
    + \frac{B_\phi}{\pom}\nabla(\rho_\rmp\pom\hat{B}_\phi),
\eeq
using standard vector
identities. To simplify the third term on the RHS
of \eqref{2and3RHS} we use
\beq
\bB\times\be_\phi=\bB_{pol}\times\be_\phi=-\frac{\nabla u}{\pom}.
\eeq
Now, putting these two results into the expression for the magnetic
force \eqref{2and3RHS} we find that:
\beq
-\frac{4\pi}{h_c}\scl 
   = \rho_\rmp\nabla B
        +\frac{1}{\pom^2}\nabla u\times
                      \nabla(\rho_\rmp\pom\hat{B}_\phi)
        +\frac{B_\phi}{\pom}\nabla(\rho_\rmp\pom\hat{B}_\phi)
        +\brac{\frac{\nabla\rho_\rmp\cdot\nabla u}{\pom B}
                     -\rho_\rmp\hat\j_\phi}\frac{\nabla u}{\pom}.
\eeq
By axisymmetry, the toroidal component of $\scl$ must be zero. Now,
all of the terms in the above expression are gradients of scalars (and
hence poloidal), except the cross product. This term
is purely toroidal and hence must be zero:
\beq \label{parallelnablas}
\frac{1}{\pom^2}\nabla u\times\nabla(\rho_\rmp\pom\hat{B}_\phi)=0.
\eeq
Hence we can remove this term from the magnetic force to arrive at a
somewhat simplified result, in terms of the
gradients of various scalar functions:
\beq \label{fmag_general}
-\frac{4\pi}{h_c}\scl 
   = -\frac{4\pi}{h_c}\rho_\rmp\nabla M
   = \rho_\rmp\nabla B
        +\frac{B_\phi}{\pom}\nabla(\rho_\rmp\pom\hat{B}_\phi)
        +\brac{\frac{\nabla\rho_\rmp\cdot\nabla u}{\pom B}
                     -\rho_\rmp\hat\j_\phi}\frac{\nabla u}{\pom},
\eeq
where $M$ is defined by equation \eqref{M_defn}, as before. For
comparison, the equivalent form for the magnetic force in the
normal-MHD derivation (see, e.g., the appendix of \citet{1f_eqm}) is:
\beq \label{lor_vs_fmag}
-4\pi\rho\nabla M = \frac{B_\phi}{\pom}\nabla(\pom B_\phi)
                     - 4\pi j_\phi \frac{\nabla u}{\pom}
\eeq
where $j_\phi=\frac{1}{4\pi}[\curl\bB]_\phi$. We see that in the
superconducting case there are extra terms with $\nabla B$ and
$\nabla\rho_\rmp$ factors.

The above derivation is similar in approach to that for the
Grad-Shafranov equation of barotropic normal MHD \citep{1f_eqm}, up
until the result \eqref{fmag_general}. One remaining step from the normal-MHD
derivation cannot, however, be generalised straightforwardly for
superconducting matter: we no longer have $\bB\cdot\nabla M=0$ and so
in general $M\neq M(u)$. This prevents us from continuing the derivation for
poloidal/mixed fields in the normal-MHD manner. Rather than discussing
this issue in more detail now, we 
stop at the interim general result 
\eqref{fmag_general} and only consider the simplest specific case,
where the magnetic field is purely toroidal. We intend to return to
the cases of purely poloidal and mixed poloidal-toroidal fields in
future work.

\subsection{Purely toroidal fields}

For purely toroidal fields, let us return to equation
\eqref{fmag_general}. In this case we have $\nabla u=0$ and the
expression for the magnetic force reduces to
\beq \label{fmag_tor}
\scl = \rho_\rmp\nabla M
     = -\frac{h_c}{4\pi}\frac{1}{\pom} \nabla(\pom\rho_\rmp B_\phi),
\eeq
where we have also used $\hat{B}_\phi=1$. Again, we recall the
normal-MHD result \citep{1f_eqm} for comparison:
\beq
\lor = \rho\nabla M
     = -\frac{B_\phi}{4\pi\pom}\nabla(\pom B_\phi).
\eeq
Now, dividing equation \eqref{fmag_tor} by $\rho_\rmp$ and taking the
curl of it yields
\beq
0 = \nabla\brac{\frac{1}{\pom\rho_\rmp}}
                    \times\nabla(\pom\rho_\rmp B_\phi)
   = -\frac{1}{\pom^2\rho_\rmp^2}
            \nabla(\pom\rho_\rmp)\times\nabla(\pom\rho_\rmp B_\phi).
\eeq
Wherever the magnetic field is non-zero $\pom\rho_\rmp\neq 0$ too,
since this is the geometry of a toroidal field, so we have
\beq
0 = \nabla(\pom\rho_\rmp)\times\nabla(\pom\rho_\rmp B_\phi)
\eeq
and hence the two arguments of the gradient operators are related by some
function $\eta$:
\beq
\eta(\pom\rho_\rmp) = \pom\rho_\rmp B_\phi.
\eeq
Putting this into \eqref{fmag_tor} and defining
$\zeta\equiv\pom\rho_\rmp$ we get 
\beq \label{nabM_tor}
\nabla M = -\frac{h_c}{4\pi}\frac{1}{\zeta}\nabla\eta(\zeta)
               = -\frac{h_c}{4\pi}\frac{1}{\zeta}\td{\eta}{\zeta}\nabla\zeta
\eeq
where the corresponding magnetic field is
\beq
\bB=B_\phi\be_\phi=\frac{\eta(\zeta)}{\zeta}\be_\phi.
\eeq
The first integral of \eqref{nabM_tor} gives our final result, which
may be used in the code:
\beq
M = -\frac{h_c}{4\pi}
             \int_0^{\pom\rho_\rmp} \frac{1}{\zeta}\td{\eta}{\zeta}\ \rmd\zeta.
\eeq
Not all functional forms of $\eta(\zeta)$ are acceptable choices ---
for example, taking $\eta$ to be a linear function of $\zeta$ leads to a
magnetic force which diverges on the polar axis. In this paper we will 
work with two other choices of $\eta$. The first (which we will
refer to as `$\zeta^2$-superconductivity' for brevity) is
\beq
\eta=\eta_0\zeta^2
\eeq
with $\eta_0$ a constant, which may be varied to adjust the magnetic
field strength. With this we have
\beq
M = -\frac{h_c\eta_0}{2\pi}\pom\rho_\rmp
\ \ \ \textrm{and}\ \ \ 
B_\phi = \eta_0\pom\rho_\rmp.
\eeq
This form of $B_\phi$ is the same as that in our normal-MHD
toroidal-field case. \skl{Note that 
  $\zeta^2$-superconductivity is the two-fluid equivalent of the case
  considered by \citet{akgun_wass}.}
The second functional form we
will work with (hereafter `$\zeta^3$-superconductivity') is
\beq
\eta=\eta_0\zeta^3.
\eeq
The magnetic force scalar and magnetic field in this case are given by
\beq
M = -\frac{3h_c\eta_0}{8\pi}\pom^2\rho_\rmp^2
\ \ \ \textrm{and}\ \ \ 
B_\phi = \eta_0\pom^2\rho_\rmp^2.
\eeq
\skl{At this stage, our motivation for choosing these two functional forms
is their simplicity. Later, we will find that the resulting equilibria
in the two cases are very similar, suggesting that our results may be
quite generic.}

\section{Numerics}

\subsection{Overview of code}

The code we use iteratively solves the equilibrium equations for our neutron
star model, and being non-linear is not restricted to the perturbative
regime of slow rotation and weak magnetic fields. Our numerical scheme
is based on the Hachisu  
self-consistent field (SCF) method \citep{hachisu}, a more robust
extension of an earlier SCF method by \citet{ostr_mark}. A version of
the Hachisu SCF method for magnetised stars was presented by
\citet{tomi_eri}. Since our scheme is a fairly straightforward extension
of these previous ones, we content ourselves with a summary here and
focus on the differences between them; we refer the reader to
\citet{hachisu} and \citet{tomi_eri} for more details.

To find equilibrium models, the user must specify a number of
stellar parameters at the outset. The major ones are related to:\\
1. the EOS --- through the polytropic indices $N_\rmn$ and $N_\rmp$ (related to the
compressibility of each fluid);\\
2. the shape of the star (related to the rotation rate) --- specified
through the ratio of polar to equatorial radii $r_{pole}/r_{eq}$, and the
ratio of the neutron-fluid equatorial surface to the proton-fluid
equatorial surface $r_{eq}^\rmn/r_{eq}^\rmp$;\\
3. the magnetic field --- through $\kappa$ and $a$ for normal-MHD
mixed fields (or poloidal fields), $\lambda$ for normal-MHD toroidal
fields, and $\eta_0$ for toroidal fields in a superconductor. See
sections 2.2 and 3.2 for more details.

We nondimensionalise all quantities within the code using the
requisite combination of $r_{eq}$, $G$ and $\rhom$ and use a hat
to denote these dimensionless quantities. As a consequence we have
$\hat{r}_{eq}=\hat\rho_{\max}=1$. When calculating integrals we first
decompose the integrands into radial pieces and Legendre polynomials
(note that there is no azimuthal piece, since we work in
axisymmetry). The contributions over all relevant grid points are then
summed up appropriately. We include angular contributions up to degree
$l=32$ in the Legendre polynomials, thus allowing for magnetic
configurations of high multipolar structure, in
contrast with the dipolar models of \citet{GAL}.

\subsection{Finding integration constants and the rotation rate}

To find the stellar rotation rate and integration constants of the
Euler equations we have to evaluate these equations at suitable
boundaries. Typically the two fluid surfaces do not coincide, and
there are different procedures to calculate these constants depending
on whether the proton or neutron fluid is outermost. However, we know
\emph{a posteriori} that all our configurations have the proton fluid
either outermost or coincident with the neutron surface, so we only
consider that case here.

For this, we work with the proton-fluid Euler and the difference Euler
and aim to find $C_\rmp,C_d$ and $\Omega$; recall that we
assume that both fluids have the same rotation rate,
$\Omega_\rmn=\Omega_\rmp=\Omega$. In this section we define each fluid
surface by the vanishing of its corresponding chemical potential
$\tilde\mu_\rmx=0$; this approach was also used by \citet{yosh_eri}
and avoids numerical difficulties associated with dealing with the
true fluid surfaces --- those given by $\rho_\rmx=0$. In general the
two definitions do not agree, due to coupling between the different
fluids \citep{prix_nc}, and in practice one would expect an interior phase
transition rather than a sharp fluid surface. Despite these issues,
working with the $\tilde\mu_\rmx=0$ surfaces is sufficient for our
simplified model of a two-fluid NS. In section 5.2 we describe how to
check the surfaces of the redimensionalised, physical, star.

This subsection requires the user-specified stellar axis ratio
$r_{pole}/r_{eq}$ (which is the same as the proton-fluid axis ratio
$r_{pole}^\rmp/r_{eq}^\rmp$). In our dimensionless units the axis ratio is
equal to the dimensionless polar radius of the proton fluid
$\hat{r}_{pole}^\rmp$. The neutron-surface equatorial radius
$\hat{r}_{eq}^\rmn=r_{eq}^\rmn/r_{eq}^\rmp$ is also used here. 
We begin by evaluating \eqref{pEuler} at the equatorial stellar
surface $\hat{r}_{eq}=1$:
\beq
\hat{\mu}_\rmp(1)=0
   =\hat{C}_\rmp+\hat{M}(1)+\frac{\hat\Omega^2}{2}-\hat\Phi(1).
\eeq
and at the polar stellar surface $\hat{r}_{pole}(=\hat{r}_{pole}^\rmp)$:
\beq
\hat{\mu}_\rmp(\hat{r}_{pole})=0
   =\hat{C}_\rmp+\hat{M}(\hat{r}_\rmp)-\hat\Phi(\hat{r}_\rmp).
\eeq
We may combine the above expressions to find $\Omega$ and $C_\rmp$:
\beq
\hat\Omega^2
  =2\Big( \hat{M}(\hat{r}_{pole})-\hat{M}(1)
        -\hat\Phi(\hat{r}_{pole})+\hat\Phi(1) \Big)
\eeq
\beq
\hat{C}_\rmp=\hat\Phi(1)-\frac{\hat\Omega^2}{2}-\hat{M}(1).
\eeq
Note that \skl{we have} $M\propto\rho_\rmp$ for a toroidal field (in
both normal and
superconducting matter) and so is zero at
the stellar surface; in this case the expressions for $\Omega$ and
$C_\rmp$ have no explicit dependence on the magnetic force.
Now evaluating the difference Euler at the \emph{neutron-fluid surface},
rather than the stellar surface, we determine $C_d$:
\beq
\hat{C}_d = \hat\mu_\rmp\brac{\hat{r}_{eq}^\rmn}
             - \hat{M}\brac{\hat{r}_{eq}^\rmn}.
\eeq

\subsection{Iterative step}

We recall that the chemical potentials are defined from the energy
functional $\clE$ by
\beq
\tilde\mu_\rmx\equiv\left.\pd{\clE}{\rho_\rmx}\right|_{\rho_{\mathrm{y}}}.
\eeq
For our particular functional \eqref{eos_functional}, the resulting
expressions for the chemical potentials may be rearranged to give
\beq \label{otherinvert}
\rho_\rmn = \brac{\frac{\mun}{k_\rmn\gamma_\rmn}}^{N_\rmn}\ ,\ 
\rho_\rmp = \brac{\frac{\mup}{k_\rmp\gamma_\rmp}}^{N_\rmp}.
\eeq
Evaluating these at the centre of
the star we obtain relations between the maximum values of the 
chemical potential and density for each fluid. In dimensionless form
these may be rearranged to show that
\beq
\hat{k}_\rmp\gamma_\rmp=\hat\mu_\rmp^{\max} x_\rmp(0)^{-1/N_\rmp}\ ,\ 
\hat{k}_\rmn\gamma_\rmn=\hat\mu_\rmn^{\max}(1-x_\rmp(0))^{-1/N_\rmn},
\eeq
where $x_\rmp(0)$ is the central proton fraction, using the fact that
$\hat\rho_\rmp=\rho_\rmp/\rho=x_\rmp$ and similarly
$\hat\rho_\rmn=(1-x_\rmp)$.
\skl{Comparing these relations with those in \eqref{otherinvert} we obtain
expressions which will allow us to iterate for new density
distributions within our numerical scheme:}
\beq \label{it_rhopn}
\hat{\rho}_\rmp 
   = x_\rmp(0)
      \brac{\frac{\hat{\mu}_\rmp}{\hat{\mu}_\rmp^{\max}}}^{N_\rmp}
\ \ ,\ \ 
\hat{\rho}_\rmn 
   = \big(1-x_\rmp(0)\big)
      \brac{\frac{\hat{\mu}_\rmn}{\hat{\mu}_\rmn^{\max}}}^{N_\rmn}.
\eeq

\subsection{Iterative procedure}

We now have all equations needed for the iterative procedure that
generates our equilibria. As discussed at the start of this section,
the procedure is an extension of the Hachisu SCF method
\citep{hachisu,tomi_eri}. \skl{To begin the iterative process (step 0),
we set the two fluid densities and the magnetic potential to be
constant within the star; however, we have found that the final
equilibria are independent of the values chosen at this step.} Note
that for a toroidal field --- in  
normal or superconducting matter --- there is no separate iteration for
the magnetic field and hence no step 2; instead the behaviour of the
field is directly linked to $\rho_\rmp$. The iterative steps are as
follows:\\
 \\
0. For the initial iteration, start with a guess that
$\rho_\rmn,\rho_\rmp$ and $\Aph$ are all constant;\\
1. Calculate the gravitational potential $\Phi$ from the $\rho_\rmn$ and
$\rho_\rmp$ distributions and Poisson's equation \eqref{int_Poisson};\\
2. Calculate the new magnetic potential component $\Aph$ from its
value at the previous iteration $\Aph^{old}$, using the magnetic
Poisson equation \eqref{int_Aph} with $\Aph^{old}$ and $\rho_\rmp$ in the
integrand;\\
3. Evaluate the proton-fluid Euler \eqref{pEuler} at the equatorial
and polar surface, using boundary conditions on $\mup$; this gives two
equations which fix $\Omega^2$ and then $C_\rmp$;\\
4. We are now able to use the proton-fluid Euler to find $\mup$
throughout the star;\\
5. Evaluate the difference-Euler \eqref{dEuler} at the equatorial
neutron-fluid surface to find $C_d$;\\
6. Now use the difference-Euler to find $\mun$ throughout the star;\\
7. Calculate the new density distributions from the chemical
potentials, using the expressions in \eqref{it_rhopn};\\
8. Return to step 1 using the new $\rho_\rmn,\rho_\rmp$ and $\Aph$;
repeat procedure until satisfactory convergence is achieved
(i.e. until the difference between quantities at consecutive iterative
steps is less than some small tolerance).

\section{Stellar parameters and sequences}

\subsection{Redimensionalising}

When looking at sequences of configurations, to study the effect of
magnetic fields or rotation, we need to ensure we are always working
with the same physical star --- the dimensionless quantities produced
by the code are not sufficient. Two stars are physically the same if
they have the same mass and equation of state:
$\clM,\gamma_\rmn,\gamma_\rmp,k_\rmn$ and $k_\rmp$ must all be equal for
both. Since we specify the (dimensionless) indices $\gamma_\rmn$ and
$\gamma_\rmp$ at the outset, we only need to fix the physical values
of $\clM,k_\rmn$ and $k_\rmp$.

Fixing the stellar mass is straightforward --- we choose
$\clM=1.4\clM_\odot$ (where $\clM_\odot$ is one solar mass) --- but
more care is needed with the polytropic constants. We choose to fix 
their physical values in the following way: take a spherical star in
hydrostatic equilibrium (nonrotating and with no magnetic field), with some
fixed $\gamma_\rmn$ and $\gamma_\rmp$ and coincident neutron and
proton surfaces. Now find the polytropic constants
$k_\rmn,k_\rmp$ that give it a radius of 10 km. More specifically,
the dimensionless mass and polytropic constants are related to their
physical counterparts through:
\beq
\hat{\clM} = \frac{\clM}{\rhom r_{eq}^3}\ \ ,\ \ 
\hat{k}_\rmx=\frac{k_\rmx}{Gr_{eq}^2\rhom^{\gamma_\rmx-2}}.
\eeq
Having specifying a spherical configuration with some particular
$\gamma_\rmx$, the code may be used to calculate $\hat{\clM}$ and
$\hat{k}_\rmx$. \skl{The requirement that $\clM=1.4\clM_\odot$ and
$r_{eq}=10$ km for the spherical configuration then fixes the physical
value of $k_\rmx$ for the stellar sequence; from this we may find
$\rhom$ and $r_{eq}$ for any star in the sequence, and hence
redimensionalise any other quantities (e.g. rotation rates and field 
strengths). Any rotating, magnetised configuration redimensionalised in this
manner will be the same physical star: if you stopped it from rotating and
removed its magnetic field, it would return to the same 10 km
spherical body.}

\subsection{Adjusting the position of the fluid surfaces}
 
Our iterative procedure does not automatically adjust the
location of the neutron surface, even though it will change with
respect to the proton surface when the star is rotating or
magnetised. A rotating or magnetised star in which the two fluids
coincide at the equator would \emph{not} return to our canonical
unmagnetised nonrotating spherical star with coincident fluid
surfaces. This means that the physical values of the polytropic
constants $k_\rmx$ would be different in the two cases. Instead
we have to manually adjust the location of the neutron-fluid
equatorial surface so that the values of $k_\rmx$ \emph{do} agree with those
of our canonical physical star. In practice the difference between the
fluid surfaces is only significant at high rotation rates or magnetic
field strengths, and for the purposes of this study a simple 
root-finder algorithm can be employed to run the code a few times,
adjusting the neutron-surface until satisfactory results are achieved.

\subsection{Virial test}

The scalar virial theorem, which is derived from the (sum) Euler
equation, states that a certain combination of energies balances the
acceleration of the system. For a single-fluid polytrope:
\beq
2T+\emag+W+3(\gamma-1)U=\frac{1}{2}\td{^2 I}{t^2}
\eeq
where $T,\emag,W$ and $U$ are, respectively, the kinetic, magnetic,
gravitational and internal energies of the system, and $I$ is the
moment of inertia. For our `separable' equation of state, where each
fluid obeys its own polytropic relation, this relation may be
generalised simply:
\beq
2T+\emag+W+3(\gamma_\rmn-1)U_\rmn+3(\gamma_\rmp-1)U_\rmn
  = \frac{1}{2}\td{^2 I}{t^2}.
\eeq
Since we seek equilibrium configurations, we can test the accuracy of
our code by how close the residual acceleration of the system
(i.e. the right-hand side of the previous equation) is to zero. We
then form a dimensionless 
quantity by normalising this acceleration through division by $W$:
\beq \label{VT}
\textrm{`virial\ test'}\equiv \textrm{error} = \frac{|2T+\emag+W+3(\gamma_\rmn-1)U_\rmn+3(\gamma_\rmp-1)U_\rmp|}{|W|}.
\eeq

\subsection{Ellipticity}

Our numerical scheme involves specifying some axis ratio
$r_{pole}/r_{eq}$, corresponding to the surface shape of the
star. A more informative quantity however is the ellipticity, which measures
the deviation of the whole mass distribution from sphericity. Starting
from the mass quadrupole moment tensor
\beq
Q_{jk} = \int\rho x_j x_k \ \rmd V
\eeq
where $x_j,x_k$ are Cartesian coordinates, we define the ellipticity
of the star through the components of the quadrupole moment
at the equator $Q_{xx}$ and pole $Q_{zz}$:
\beq
\epsilon=\frac{Q_{xx}-Q_{zz}}{Q_{xx}}.
\eeq

\subsection{Magnetic-field quantities}

The conventional definition of the (normal-MHD) magnetic energy is as an
integral out to infinite radius:
\beq
\emag=\frac{1}{8\pi}\!\!\int\limits_{\mathrm{all\ space}}\!\! B^2\ \rmd V,
\eeq
which would have to be truncated on our finite numerical grid. One
alternative but equivalent form of $\emag$ is as the integral of 
$\br\cdot\lor$; this may be seen from the relevant part of the virial
theorem derivation \citep{chand_hydro}. Since $\lor$ is only non-zero
over the star (by virtue of its $\rho_\rmp$ factor), we may write the
magnetic energy as
\beq
\emag=\int\limits_{\mathrm{star}}\br\cdot\lor\ \rmd V.
\eeq
The magnetic energy in the toroidal component of the field is also an
integral over the star:
\beq
\etor=\frac{1}{8\pi}\int\limits_{\mathrm{star}}B_\phi^2\ \rmd V,
\eeq
since $B_\phi=0$ outside the star. Finally, it is clear that the
poloidal-field contribution to the energy is simply $\emag-\etor$,
so there is no need to use an integral out to infinity for this case
either.

When looking at mixed-field configurations, we want some measure of the
relative strength of the toroidal and poloidal field components. We
do this in two ways: one related to the global contributions of
each component and one which looks at their maximum values. For the
former, we take the percentage of the magnetic energy stored in the
toroidal component $\etor/\emag$; this is relevant for understanding
quantities like the ellipticity, which scale with $B^2$. For the
latter we compare the maximum value attained by each field component,
using the ratio
\beq \label{poltor_max}
\frac{\Btor^{\max}}{\Btor^{\max}+\Bpol^{\max}}.
\eeq
Unlike the more obvious choice of $\Btor^{\max}/\Bpol^{\max}$, the
ratio \eqref{poltor_max} varies from zero (no toroidal field) to unity
(purely toroidal field), which is perhaps clearer.

Although a number of studies have used the energy ratio $\etor/\emag$
to assess the stability of a magnetic-field configuration (see, e.g.,
\citet{braithtorpol}), this is a 
global quantity, whereas the relevant instability is highly localised ---
around the neutral point for a purely poloidal field
\citep{markey}. To remove this instability, what is more important is 
whether the toroidal component is \emph{locally} comparable in
strength with the poloidal one \citep{wright}. Since the strength of the poloidal
component in the closed-field line region is directly related to its
maximum value, we argue that the ratio \eqref{poltor_max}
may be a better indication of the stability of a configuration.

Since we are interested in the effect of a magnetic field on the
global properties of neutron stars, we work with a
volume-averaged form of the magnetic field $\Bav$ rather than a surface
value:
\beq
\Bav^2\equiv\frac{1}{V}
      \!\!\int\limits_{\mathrm{all\ space}}\!\!
               B^2\ \rmd V=\frac{8\pi\emag}{V},
\eeq
where $V$ is the star's volume. This volume-averaged value is
approximately double the polar surface value $B_{pole}$ for a poloidal
field (for a toroidal field the surface value is zero).

For superconducting configurations we have to fix the value of the
first critical field strength $H_{c1}$. This is not a constant, but is
related to $\rho_\rmp$ and hence position-dependent. Unless stated
otherwise, we fix the central (i.e. maximum) value of the field to be
$H_{c1}(0)=10^{16}$ G, to fit with the estimate from
\citet{GAS2011}. This corresponds to a volume-averaged value of
$\bar{H}_{c1}\approx 3\times 10^{15}$ G.

\section{Results}

Our results are divided into four subsections. We begin by testing the
convergence properties of the code and comparing our results with
those of previous studies. We then present results for two-fluid stars
in normal MHD, 
focussing on the effect of stratification. In the third subsection we
study configurations where the protons form a type-II superconductor,
and in the fourth subsection we give approximate relations for the magnetic
ellipticity of stars as a function of field strength.

\subsection{Tests of the code}

\begin{figure}
\begin{center}
\begin{minipage}[c]{0.5\linewidth}
\psfrag{virial}{virial test}
\psfrag{resolution}{grid points}
\includegraphics[width=\linewidth]{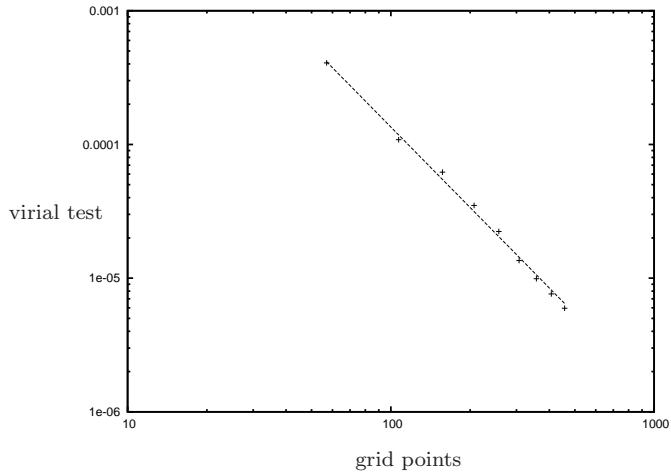}
\end{minipage}
\caption{\label{cvg_test}
         Testing the order of convergence for our code, by plotting the
         virial test result against the number of grid points. All
         results are for a rotating magnetised NS model, stratified
         with $N_\rmn=1$ and $N_\rmp=2$, with $\Bav=1.2\times 10^{17}$
         G and $\Omega=650$ Hz. The line drawn shows the expected
         result for second-order convergence; it agrees well with the
         actual points from different resolutions.}
\end{center}
\end{figure}

Before we present results for our NS models, we test how accurate our
code is by using the virial test \eqref{VT} to determine the relative
error in calculating an equilibrium configuration for different
numbers of grid points; see figure \ref{cvg_test}. We see that the
error decreases quadratically with resolution, and hence is
second-order convergent, as intended.

Next, we compare our results with the three nonrotating and
unmagnetised two-fluid
models presented in \citet{prix_r}. To do this we need to convert from
their dimensionless quantities to ours, using the speed of light
$c=3.00\times 10^{10}$ cm s${}^{-1}$ and
nuclear density $\rho_{nuc}=1.66\times 10^{14}$ g cm${}^{-3}$. Some
algebra shows that our dimensionless polytropic constants (denoted by
the index $LAG$) and those of \citet{prix_r} (with index $PR$) are related by
\beq
\hat{k}_\rmx^{LAG}
  = \frac{c^2}{Gr_{eq}^2\rho_{nuc}}
      \brac{\frac{\rho_{\max}}{\rho_{nuc}}}^{\gamma_\rmx-2}
       \ \hat{k}_\rmx^{PR}
  = 8.11\times 10^3 \brac{\frac{r_{eq}}{\mathrm{km}}}^{-2}
      \brac{\frac{\rho_{\max}}{\rho_{nuc}}}^{\gamma_\rmx-2}
       \ \hat{k}_\rmx^{PR}.
\eeq
Next we consider the conversion of dimensionless masses:
\beq
\hat{M}^{LAG} = 12000 \brac{\frac{\rho_{\max}}{\rho_{nuc}}}^{-1}
                     \brac{\frac{r_{eq}}{\mathrm{km}}}^{-3} 
                     \ \hat{M}^{PR}.
\eeq
Now using these, we can make a direct comparison between our results
and those of \citet{prix_r} --- see table \ref{PR_compare}.

\begin{table}
\begin{center}
\caption{\label{PR_compare}
         Comparing with the unmagnetised two-fluid models presented in
         Prix and Rieutord. The values shown for the polytropic constants
         and masses are our 
         results, in our dimensionless units, with the discrepancy
         from those of Prix and Rieutord indicated in brackets.}
\begin{tabular}{cccccc}
\hline
  & $\gamma_\rmn$ & $\gamma_\rmp$ & $\hat{k}_\rmn$ & $\hat{k}_\rmp$ & $\hat{M}$\\
\hline
Model I   & 2.0 & 2.0 & 0.7074 (0.1\%) & 6.366 (0.1\%) & 1.273 (0.5\%) \\
Model II  & 2.5 & 2.1 & 0.7060 (0.5\%) & 9.034 (1.4\%) & 1.834 (0.7\%) \\
Model III & 1.9 & 1.7 & 0.6802 (0.7\%) & 3.466 (0.9\%) & 1.083 (1.0\%) \\
\hline
\end{tabular}\\
\end{center}
\end{table}

\subsection{Two-fluid equilibrium configurations in normal MHD}

Our two-fluid formalism gives us a great deal of flexibility, but for
clarity we will present results using just a few canonical
values. For unstratified models, we set $N_\rmn=N_\rmp=1$; whilst when
we refer to stratified models we have taken $N_\rmn=1$ and $N_\rmp=2$
unless otherwise stated \skl{(see the following figure for
  justification of this choice)}. In addition, we have fixed the
central value of the proton fraction $x_\rmp(0)$ to be equal to $0.15$
for all results, \skl{although we have found that our results are
  virtually independent of the value chosen}. In many figures, we show
stars whose magnetic fields are 
(probably) unphysically high --- of the order of $10^{17}$ G. This is
to emphasise effects which would be less obvious at lower values. We
do, however, discuss the scaling of our results to more typical NS
field strengths too.

In this section we present results for neutron star models composed of
superfluid neutrons and normal-fluid protons, subject to a magnetic
field. As discussed above, this situation may apply to an interior
region of magnetars, where it is conceivable that the second
critical field $H_{c2}$ will be exceeded. Below this critical field
the protons will be superconducting, and we cover this case in the
following subsection.

\begin{figure}
\begin{center}
\begin{minipage}[c]{0.9\linewidth}
\psfrag{rho_tot}{$\displaystyle{\frac{\rho_{tot}}{\rho_{\max}}}$}
\psfrag{r_eq}{$\ r$}
\psfrag{x_p}{$x_\rmp$}
\psfrag{N_p=1}{$N_\rmp=1$}
\psfrag{N_p=2}{$N_\rmp=2$}
\psfrag{N_p=3}{$N_\rmp=3$}
\includegraphics[width=\linewidth]{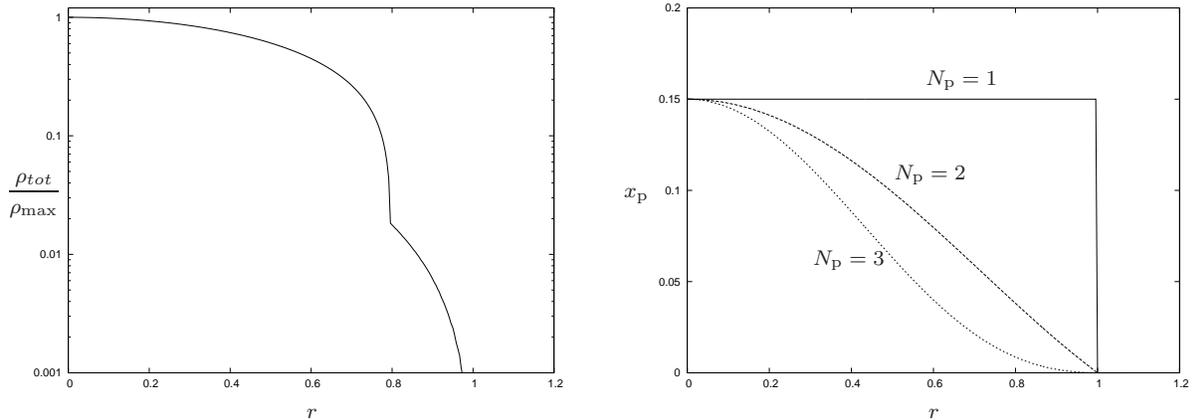}
\end{minipage}
\caption{\label{EOS_fits}
         Demonstrating the additional flexibility possible with a
         two-fluid stellar model. Left: an approximation to the
         Douchin-Haensel EOS for a fluid core and a solid crust,
         \skl{plotting the total fluid density
         $\rho_{tot}=\rho_\rmn+\rho_\rmp$ against radius $r$}. In 
         Douchin and Haensel, the core is fairly close to a polytrope with index
         $N_{\rm core}\approx 0.7$ and the crust is similar to a different
         polytrope, having $N_{\rm crust}\approx 1.5$, with a
         crust-core boundary at $\approx 0.03\rhom$ and central
         proton fraction $x_\rmp(0)=0.15$. In our model we have
         $x_\rmp(0)=0.15$ and $N_{\rm core}\approx N_\rmn=0.7$, while
         $N_{\rm crust}\approx N_\rmp=1.3$. This gives us a density of
         $\sim 0.02\rhom$ at the crust-core boundary, with a 2
         km-thick `crust' for a 10 km radius star. Right: the change in
         proton fraction $x_\rmp$ against radius. We fix
         $N_\rmn=1$ and look at the variation of proton fraction with
         $N_\rmp$. In the unstratified case, $N_\rmp=N_\rmn=1$, the
         proton fraction is constant within the star.}
\end{center}
\end{figure}

In figure \ref{EOS_fits} we show the additional flexibility possible
in two-fluid models. We can model the effect of having a fluid
core surrounded by a crust, by allowing the two fluids to terminate at
different points, so that there is an outer region with only one of
the fluid species --- in this case the protons. This is shown in the
left-hand plot, where we attempt to fit the equation of state
described in \citet{douchin} by adjusting parameters in our model
accordingly. Two-fluid models also allow us to have stratification and
consequently a non-uniform proton fraction, as one would expect in
real NSs \citep{kaminker_hy}; see the right-hand plot. Comparing
this plot with figure 1 from \citet{GAL} showing proton
fractions in more sophisticated models, we see that choosing $N_\rmp=2$
produces a reasonable-looking proton-fraction profile.

\begin{figure}
\begin{center}
\begin{minipage}[c]{0.8\linewidth}
\includegraphics[width=\linewidth]{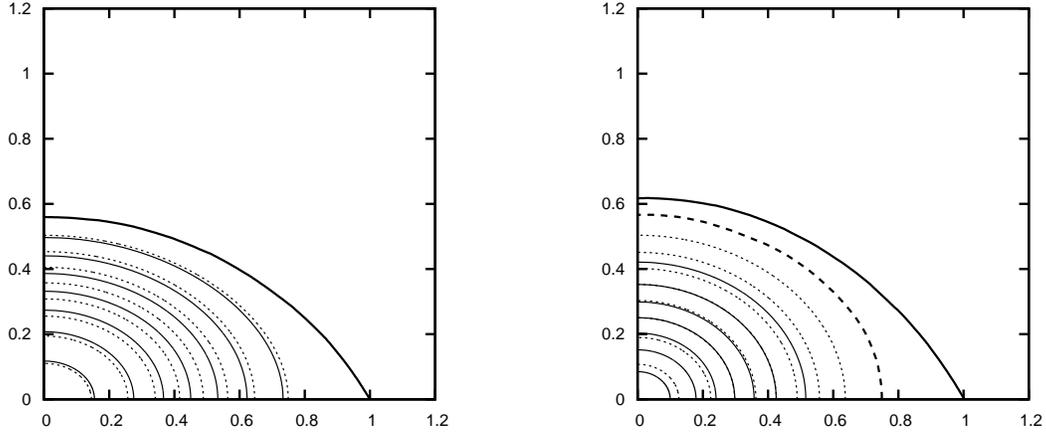}
\end{minipage}
\caption{\label{breakup}
         Density contours of two unmagnetised stellar configurations
         rotating at breakup velocity, left: our canonical
         unstratified model ($N_\rmn=N_\rmp=1$), right: our canonical
         stratified model ($N_\rmn=1$ and $N_\rmp=2$). The bold lines
         show the fluid surfaces --- which are
         coincident for the unstratified star. For the stratified star the
         neutron fluid is not rotating at breakup velocity, but the
         proton fluid is.}
\end{center}
\end{figure}

Before turning to the effect of a magnetic field on a two-fluid star, we
consider the role of rotation on it. This allows us to make contact
with models in previous work, as well as giving us some
intuition about what to expect from magnetic fields. In figure
\ref{breakup} we show two configurations rotating at their Keplerian
frequency. The left-hand star is an unstratified model
($N_\rmn=N_\rmp=1$) and both fluid 
surfaces coincide. The interior density contours are different,
however, because the central proton fraction $x_\rmp(0)=0.15$; were it
$x_\rmp(0)=0.5$ then each species would have the same density contours
too. The right-hand star is a stratified model and in this case it is
the outer fluid, the protons, that determine the star's breakup
velocity --- the neutron surface is more rounded, without the
characteristic cusped-shape of a fluid at breakup \citep{prix_nc}. If
we had $N_\rmn>N_\rmp$, the neutron fluid would be outermost; however,
we do not consider this case. Instead the protons in our models are always
outermost, for rotating configurations but also magnetised ones, as we
will see later in this section.

\begin{figure}
\begin{center}
\begin{minipage}[c]{0.8\linewidth}
\includegraphics[width=\linewidth]{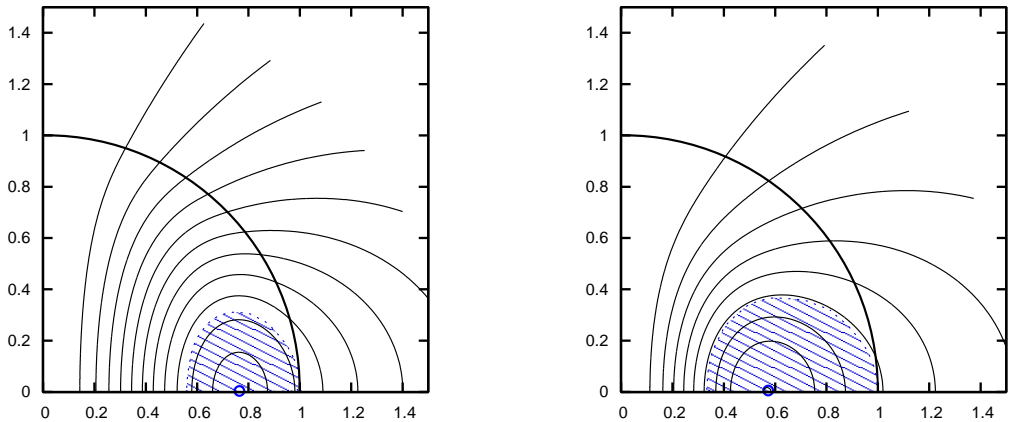}
\end{minipage}
\caption{\label{Bpol_vs_Np}
         Contours of the streamfunction $u$ for a star with a purely
         poloidal field, corresponding to magnetic field lines. The
         thick arc indicates the stellar surface, at a dimensionless
         radius of unity; the shaded area indicates the closed-field
         line region, and the small circle on the $x$-axis shows the
         location of the neutral line, where the field strength
         vanishes. Left: no stratification ($N_\rmn=N_\rmp=1$),
         right: a stratified model, with $N_\rmn=1$ and
         $N_\rmp=3$. The closed field line region is bigger in the 
         stratified case.}
\end{center}
\end{figure}

Whilst our formalism is able to deal with stars that are both highly
magnetised and rapidly rotating, we will concentrate on non-rotating
models from now on. This is because we are chiefly concerned with the
effect of stratification on magnetic fields in stars; allowing for the
extra effect of rotation obfuscates the picture. This makes our
results most directly applicable to magnetars (which rotate very
slowly), but the ellipticity formulae presented at the end of this
results section are equally applicable to the magnetic distortions in
less-magnetised NSs (like pulsars).

We begin by looking
at stars with a purely poloidal magnetic field - figure
\ref{Bpol_vs_Np}. We show contours of the streamfunction $u$, which
are parallel to magnetic field lines, shading the closed-field line
region and marking the `neutral line' (where the field vanishes) with
a circle. The field configuration of the unstratified model is very
similar to that of a single-fluid polytrope with index $N=1$ (see,
e.g., figure 3 from \cite{1f_eqm}). Stratification has the effect of
moving the neutral line inwards and increasing the volume of the
closed-field line region. Fixing $N_\rmn=1$, we find that the greater
the value of $N_\rmp$, the larger the closed-field line region; we
have emphasised the effect by taking $N_\rmp=3$ rather than our
canonical choice of $N_\rmp=2$.

\begin{figure}
\begin{center}
\begin{minipage}[c]{\linewidth}
\includegraphics[width=\linewidth]{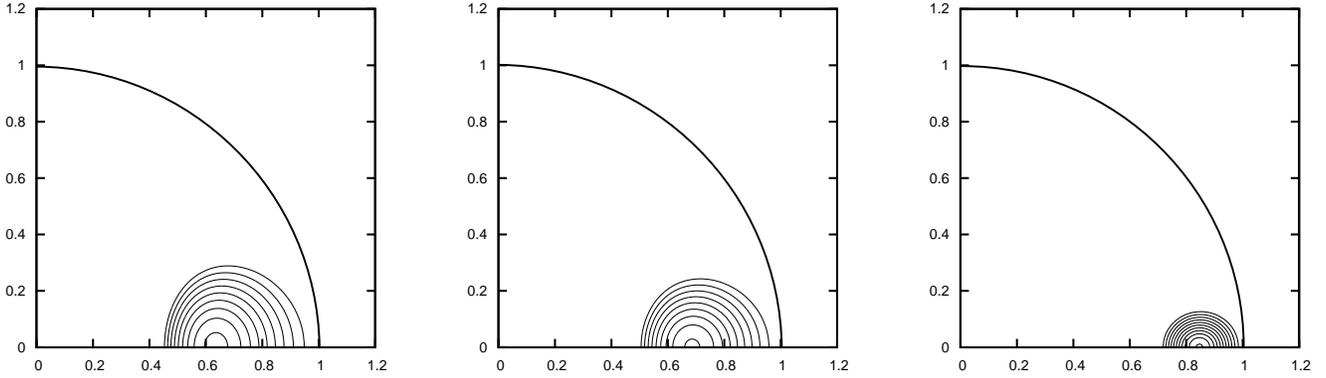}
\end{minipage}
\caption{\label{increase_a}
         Effect of increasing $a$, the coefficient of the magnetic
         function $f(u)$, within a stratified star. Contours of
         the toroidal field component are 
         shown. As $a$ is increased, from left to right, the
         percentage of toroidal field increases (1\%,3\% and 4.5\%),
         with a corresponding increase in the ratio
         $\Btor^{\max}/(\Bpol^{\max}+\Btor^{\max})$ --- from 0.10 to
         0.18, and finally 0.34. At the same time, however, the size
         of the toroidal-field region is decreased.}
\end{center}
\end{figure}

The toroidal-field component of a mixed-field star is defined by a
function $f(u)$, which can only be non-zero within the closed-field
line region, and by the coefficient $a$ (see equation
\eqref{f_defn}). Increasing $a$ increases the 
percentage of magnetic energy stored in the toroidal-field component,
and also the maximum value it attains relative to the poloidal
component, but it \emph{decreases} the volume of the torus containing the
toroidal-field component; see figure \ref{increase_a}. This seems to
limit the size of the toroidal component, as in the barotropic case
\citep{1f_eqm,ciolfi}. The volume of 
the torus is increased somewhat for higher values of $N_\rmp$, however. 

\begin{figure}
\begin{center}
\begin{minipage}[c]{0.5\linewidth}
\includegraphics[width=\linewidth]{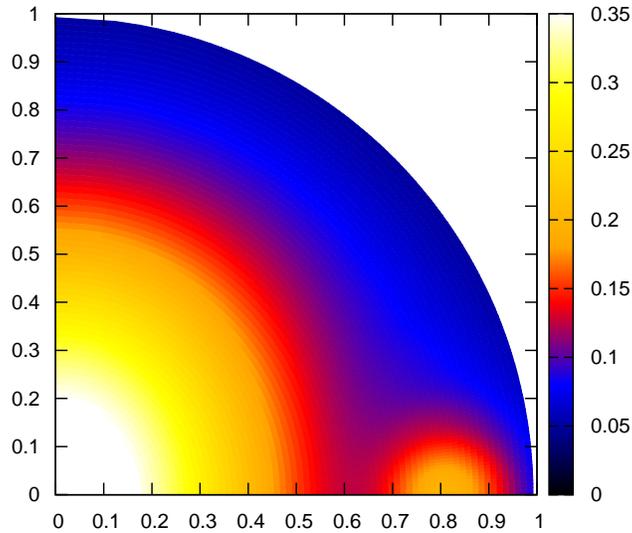}
\end{minipage}
\caption{\label{P2_Btot}
         Magnetic field strength within a stratified nonrotating NS
         with 4.7\% toroidal energy. The toroidal field is contained
         within the small region near the equatorial surface, reaching
         its maximum at $x\approx 0.8$. The poloidal field pervades the
         rest of the star, reaching a maximum at the centre. The
         surface field strength is about 20\% of the central
         value, 50\% of the maximum toroidal-component
         strength and 50\% of the average value $\Bav$.}
\end{center}
\end{figure}

In figure \ref{P2_Btot} we show the variation of magnetic field
strength within a typical mixed-field stratified configuration. The
field reaches a local maximum in two places --- the centre of the star
(corresponding to the maximum value of the poloidal component) and
near the equatorial surface (the toroidal-component maximum). The
toroidal component vanishes at the stellar surface, but the poloidal
component extends outside the star (not shown). The configuration
shown has 4.7\% toroidal energy, around the maximum value possible
within our approach. The surface field strength is about 20\% of
the central value, about 50\% of the maximum toroidal field
strength and also about 50\% of the volume-averaged field
strength $\Bav$.

\begin{figure}
\begin{center}
\begin{minipage}[c]{0.8\linewidth}
\includegraphics[width=\linewidth]{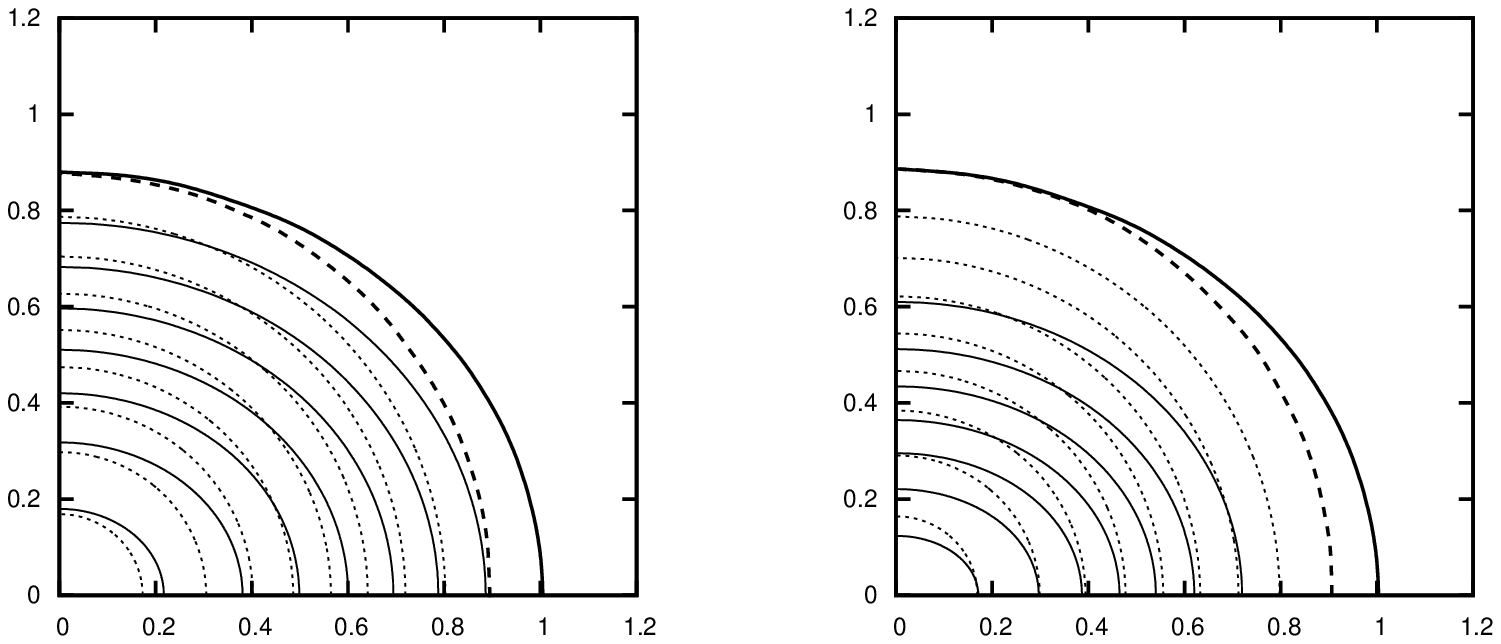}
\end{minipage}
\caption{\label{pol_densconts}
         Density contours for proton fluid (solid line) and neutron
         fluid (dashed) in non-rotating stars distorted by purely
         poloidal magnetic fields. The bold lines represent the fluid
         surfaces, which coincide at the pole but not the equator. The
         proton fluid is seen to be more distorted, due to the
         magnetic field. The left and right plots show our canonical
         unstratified and stratified models, respectively.}
\end{center}
\end{figure}

Next we turn to the effect of the magnetic field on the density
distribution of the star. We begin with purely poloidal fields, in
figure \ref{pol_densconts}.  As in
the single-fluid barotropic case, these make the star oblate, but
virtually all of the distortion is in the proton fluid, with the
neutron fluid remaining nearly spherical. The two fluid surfaces
coincide at the pole, but at the equator the Lorentz force distorts
the proton fluid, resulting in a single-fluid region composed only of
protons. To make the effect noticeable, we have shown a highly
magnetised model, with $\Bav\sim 10^{17}$ G. The single-fluid region
is analogous to the case of a rotating stratified star with no
magnetic field; see figure \ref{breakup}. As mentioned in figure
\ref{EOS_fits} and also by \citet{prix_nc}, this single-fluid region
can be thought of as an approximation to a neutron-star crust.

The unstratified (left) and stratified (right) plots of figure
\ref{pol_densconts} are very similar; the only obvious difference is that when
$N_\rmp=2$ the protons 
have a larger low density region than the $N_\rmn=1$ case (this is
true in unmagnetised stars too). We have not included density
contours for mixed-field stars, as these are very similar to those in
figure \ref{pol_densconts}, with the contour lines pushed outwards
slightly in the region where the toroidal component is located.

\begin{figure}
\begin{center}
\begin{minipage}[c]{0.4\linewidth}
\includegraphics[width=\linewidth]{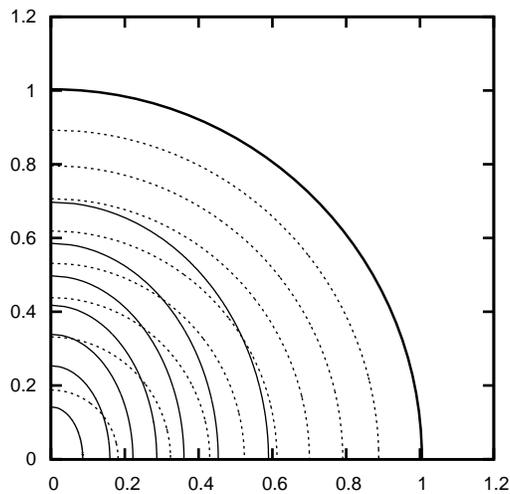}
\end{minipage}
\caption{\label{tor_densconts}
         Density contours for proton fluid (solid line) and neutron
         fluid (dashed) in a non-rotating stratified star with a purely
         toroidal magnetic field. The bold line represents the two
         (virtually coincident) fluid surfaces, and hence the stellar
         surface. Again, the proton fluid is seen to be more
         distorted, due to the magnetic field.}
\end{center}
\end{figure}

In figure \ref{tor_densconts} we show density contours of a stratified
star with a purely toroidal magnetic field. Again, the neutron fluid is
virtually spherical and unaffected by the magnetic field, whilst the
proton fluid is highly distorted --- into a prolate configuration this
time. Note that whilst the innermost proton-density contours are
noticeably prolate, the outer ones become virtually spherical; the same effect
was seen in the (single-fluid) models of \citet{ostr_hart}, who
considered mixed fields but with a dominant toroidal
component. Related to this effect, the surfaces of the two fluids are
seen to be coincident (or very close to being so). The 
unstratified case is very similar, just with more evenly-spaced
proton-density contours (as for the previous figure) --- it will be
shown in the following subsection, when we compare toroidal fields
in normal MHD and superconductivity.

\begin{figure}
\begin{center}
\begin{minipage}[c]{0.6\linewidth}
\psfrag{Np}{$N_\rmp$}
\psfrag{etor_pc}{$\displaystyle\frac{\etor}{\emag}$ (\%)}
\psfrag{Btor_max}{$\displaystyle\frac{\Btor^{\max}}{\Btor^{\max}+\Bpol^{\max}}$}
\includegraphics[width=\linewidth]{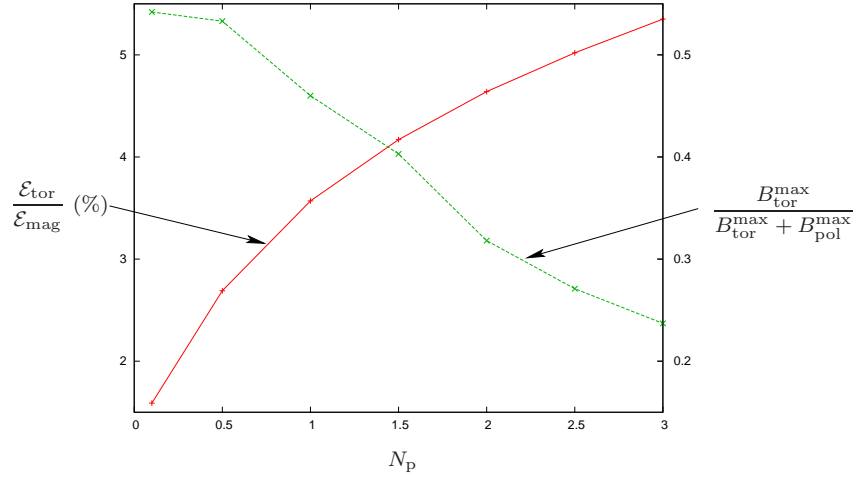}
\end{minipage}
\caption{\label{maxtor_vs_Np}
         Mixed-field configurations with the maximum possible toroidal
         component, as a function of $N_\rmp$. All configurations have
         $N_\rmn=1$, are nonrotating and have axis ratios close to
         unity. The central proton fraction $x_\rmp(0)=0.15$ in all
         cases, but the results are virtually invariant of the value
         chosen. We see that in more stratified stars the percentage
         of toroidal energy can increase, but the maximum toroidal
         field strength decreases with respect to the maximum poloidal
         field strength.}
\end{center}
\end{figure}

In figure \ref{maxtor_vs_Np} we explore mixed-field configurations
with the maximum possible toroidal component (within our models, at
least). We use two different measures of the 
relative strength of the toroidal component: comparing the maximum
values of the two field components using the ratio
$\Btor^{\max}/(\Btor^{\max}+\Bpol^{\max})$ and looking at the the
contribution of the toroidal component to the total magnetic energy,
$\etor/\emag$. Fixing $N_\rmn=1$ as usual, we
find that for increasing $N_\rmp$ the energy percentage increases, but
the relative maximum value of the toroidal component decreases. In all
cases the toroidal component is smaller --- especially in terms of
$\etor/\emag$. Note that all configurations plotted are close to
spherical (with an axis ratio of $r_{pole}/r_{eq}=0.996$). Very highly
distorted stars (with consequently strong magnetic fields) can have
slightly larger values of $\etor/\emag$.

\begin{figure}
\begin{center}
\begin{minipage}[c]{0.5\linewidth}
\psfrag{Bav}{$\displaystyle\brac{\frac{\Bav}{10^{17}\ \mathrm{G}}}^2$}
\psfrag{ellip}{$\epsilon$}
\psfrag{a}{(a)}
\psfrag{b}{(b)}
\includegraphics[width=\linewidth]{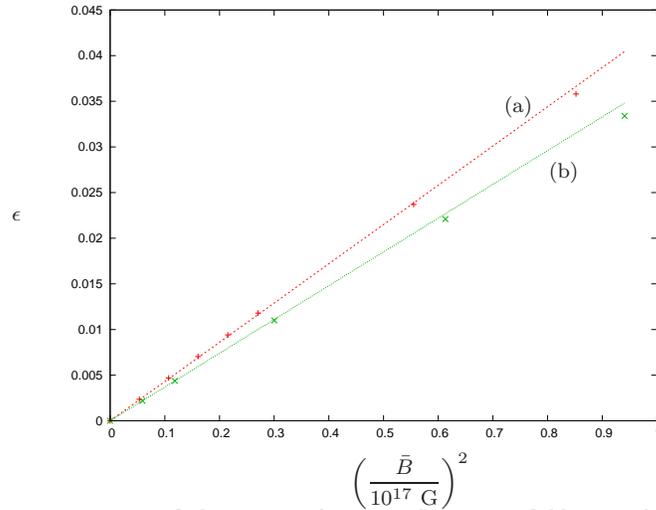}
\end{minipage}
\caption{\label{polmix_ellips}
         The ellipticity of nonrotating unstratified stars as a function of
         magnetic field strength. Line (a) shows the dependence for a
         poloidal field, whilst (b) shows a mixed-field configuration
         with 3\% toroidal energy. The toroidal component decreases
         the oblateness of the star, as expected. There is no
         discernable difference in the corresponding figure for a
         stratified star; the ellipticity appears to be virtually
         independent of stratification.}
\end{center}
\end{figure}

Finally in this section, we plot the dependence of ellipticity on
magnetic field strength, figure \ref{polmix_ellips}. The relation is
quadratic, as expected from the single-fluid case, and does not appear
to depend on the stratification of the star. We show distortions of a
purely poloidal field star, and one with a relatively strong mixed
field, 3\% toroidal energy. The toroidal component decreases the
oblateness of the star, as expected, but we are not able to generate
mixed-field configurations where the toroidal component is strong
enough to induce a \emph{prolate} distortion. We are only able to
produce prolate distortions when the magnetic field is purely
toroidal; this case is covered in the next section.

\subsection{Superconducting two-fluid equilibria with toroidal fields}

In the previous section we investigated magnetic NS models with superfluid
neutrons and normal protons. This situation
could apply in (some) magnetars, if their interior magnetic fields
exceed $H_{c2}$. Most NSs, however, are likely to be
composed predominantly of superfluid neutrons and type-II superconducting
protons. In this section we present the first NS models that account
for this, specialising to configurations with purely toroidal magnetic
fields.

\begin{figure}
\begin{center}
\begin{minipage}[c]{0.8\linewidth}
\includegraphics[width=\linewidth]{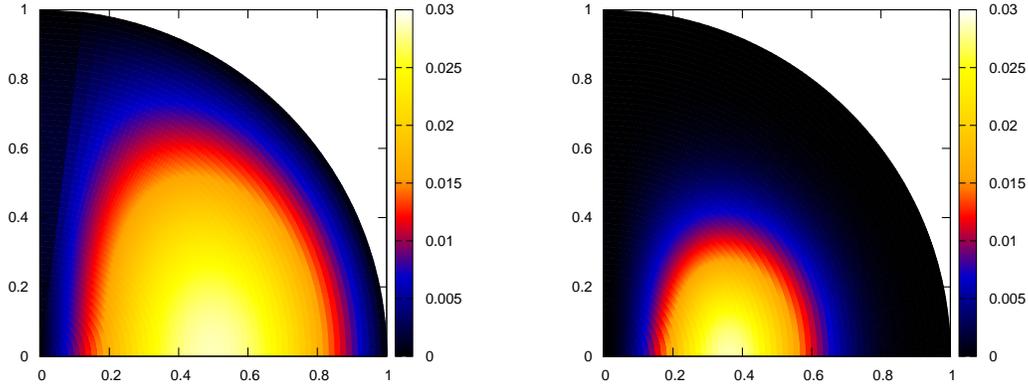}
\end{minipage}
\caption{\label{Btor_magnitude}
         The variation in field strength within stars with purely
         toroidal magnetic fields. The left-hand plot shows a
         unstratified model in normal MHD; the
         right-hand plot shows $\zeta^3$-superconductivity
         in a stratified star. Both plots have the same maximum field
         strength, for easier comparison. Stratification and superconductivity
         both have the effect of concentrating the field in a smaller,
         more central region of the star.}
\end{center}
\end{figure}

In figure \ref{Btor_magnitude} we consider the variation of
magnetic-field magnitude within the interior of two
different NS models. The left-hand plot shows a star with normal
protons and hence subject to the usual Lorentz force of MHD. It is
also unstratified. The field strength is seen to drop off fairly
slowly before vanishing at the stellar surface. In the right-hand plot
we show a toroidal magnetic field in a superconducting star, with a
particular choice of toroidal-field function which we refer to as
`$\zeta^3$-superconductivity' --- see section $3.2$. This star is also
stratified, and both this and the choice of toroidal-field function
act to `bury' the field deeper into the star. Our other choice for the
toroidal-field function, `$\zeta^2$-superconductivity', produces
configurations which are (in the unstratified case) indistinguishable
from the left-hand 
plot. This is not surprising, since the expression for the toroidal
field in this case is of the same form as for normal MHD (again, see
section $3.2$).

\begin{figure}
\begin{center}
\begin{minipage}[c]{\linewidth}
\includegraphics[width=\linewidth]{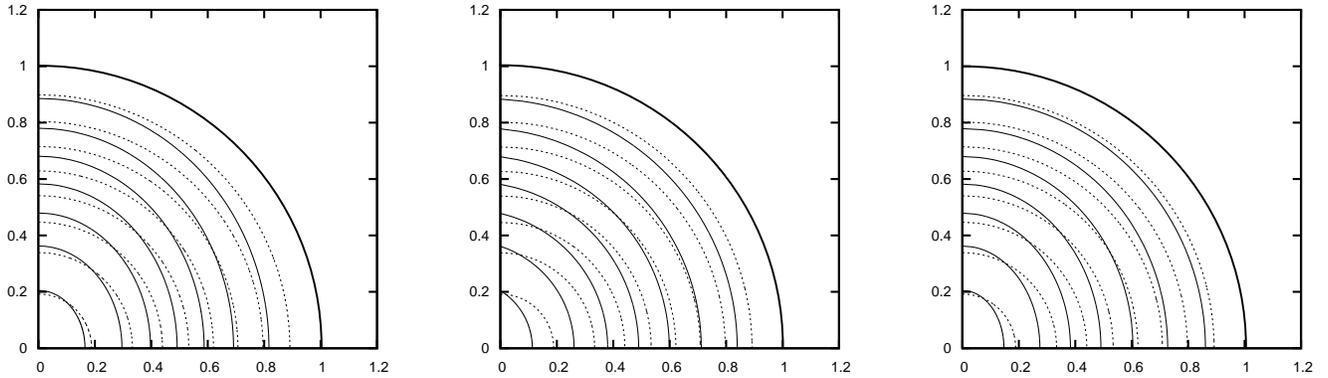}
\end{minipage}
\caption{\label{norm_vs_sc_dens}
         The distorting effect of a toroidal field. From left to right:
         normal MHD, $\zeta^2$-superconductivity,
         $\zeta^3$-superconductivity. The plots are all 
         for very high field strengths (of the order of
         $10^{17}$ G), since the aim of the figure is to illustrate
         the different ways the density distribution is distorted in
         each case. We discuss more realistic values later.}
\end{center}
\end{figure}

In figure \ref{norm_vs_sc_dens} we show density distributions for
three NS models with toroidal magnetic fields. We consider very highly
magnetised stars, which serve to emphasise the magnetic
distortions. All three models are broadly
similar: each has a spherical neutron-fluid distribution, coincident
neutron and proton-fluid surfaces, and prolate distortions of the
proton fluid. Comparing
the density contours for each model, we see that the superconducting
models have less distortion to the proton fluid in the outer region,
and more in the centre. The central panel ($\zeta^2$-superconductivity)
has the unusual feature of cusps in its proton-fluid distribution
around the pole. We show unstratified models, but the only obvious
effect of stratification is that the spacing of the proton-fluid density
contours is altered, as seen earlier in figure \ref{pol_densconts}.

We mentioned that figure \ref{norm_vs_sc_dens} shows models with
extremely strong magnetic fields, and we will show models with similar field
strengths in the next figure too. Whilst these high fields are
useful for emphasising certain features, the main reason for using
them comes from our numerical method, which calculates configurations
based on a given distortion (axis ratio). Since magnetic distortions
are typically very small, this means that even the axis ratio for the
least non-spherical star we can specify (constrained by the grid resolution)
corresponds to a very strong magnetic field. This is not necessarily a
problem for normal MHD, but superconductivity will be broken at these
field strengths, which exceed the second critical field of
$H_{c2}\approx 10^{16}$ G. We are able to produce these models because the
destruction of superconductivity is not built into them; the equations
may be solved for any field strength. Having done so, however, we need
to check that these models are consistent with our expectations for
NSs with superconducting protons. If they are, we may extrapolate our
results back to more realistic models, where $\Bav<\bar{H}_{c2}$.

\begin{figure}
\begin{center}
\begin{minipage}[c]{0.45\linewidth}
\psfrag{Bav}{$\displaystyle\brac{\frac{\Bav}{10^{17}\ \mathrm{G}}}^2$}
\psfrag{ellip}{$-\epsilon$}
\includegraphics[width=\linewidth]{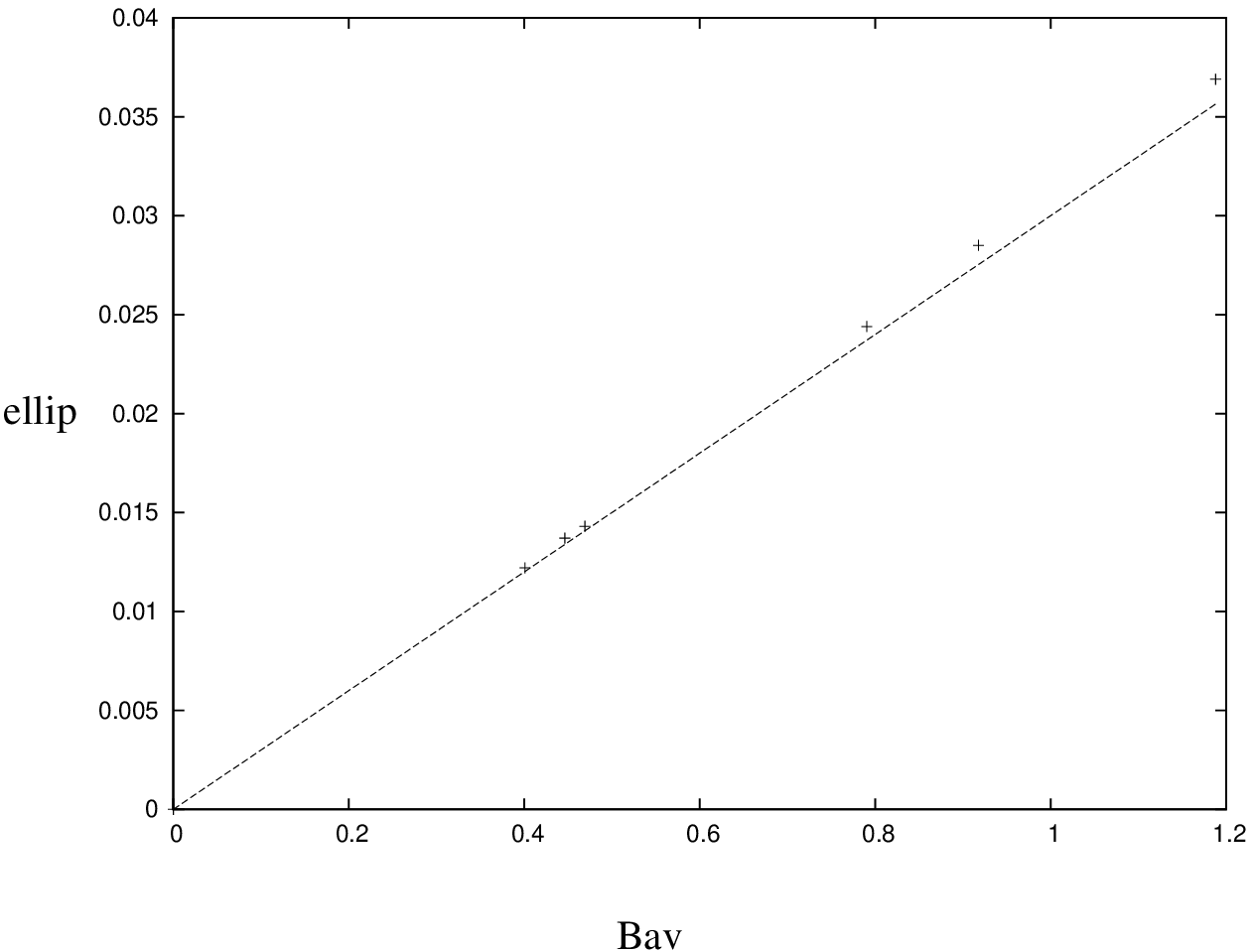}
\end{minipage}
\hspace{0.05\linewidth}
\begin{minipage}[c]{0.45\linewidth}
\psfrag{a}{(a)}
\psfrag{b}{(b)}
\psfrag{c}{(c)}
\psfrag{Bav}{$\displaystyle\frac{\Bav}{10^{17}\ \mathrm{G}}$}
\psfrag{ellip}{$-\epsilon$}
\includegraphics[width=\linewidth]{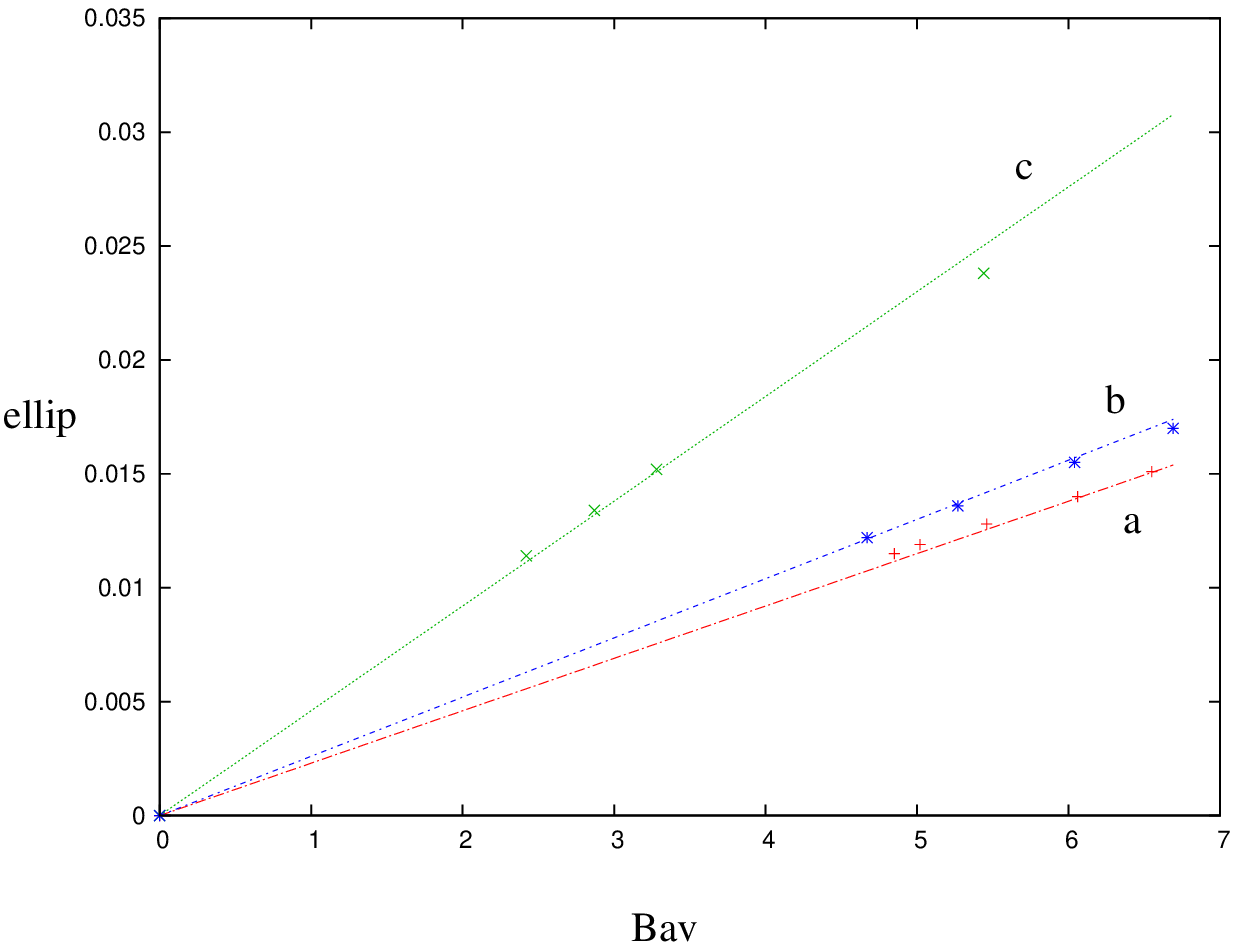}
\end{minipage}
\caption{\label{toroidal_ellips}
         Left: in normal MHD the ellipticity scales with
         $B^2$; right: in superconductivity it scales with $H_{c1}B$ - this is
         illustrated by showing the results for a central
         critical-field value $H_{c1}(0)=10^{16}$ G (a), and also
         $H_{c1}(0)=2\times 10^{16}$ G (c); the latter line has double
         the gradient. (a) and (c) are for $\zeta^2$-superconductivity; (b)
         is for $\zeta^3$-superconductivity with $H_{c1}(0)=10^{16}$ G.}
\end{center}
\end{figure}

We perform this sanity check next, in figure
\ref{toroidal_ellips}. We plot the scaling of ellipticity
with average field strength $\Bav$ for NS models with toroidal
magnetic fields, for normal (left) and type-II superconducting protons
(right). Note that all ellipticities are negative, since these
configurations are prolate. In the left-hand plot we find that the
measured numerical values agree very well with the expected scaling
$\epsilon\propto B^2$, with small deviations when $\Bav\gtrsim
10^{17}$ G.

We now turn to the right-hand plot of figure
\ref{toroidal_ellips}. This time we plot the ellipticity against
$\Bav$, not $\Bav^2$. First we look at configurations with
$\zeta^2$-superconductivity, for a central critical field value
$H_{c1}(0)=10^{16}$ G (the points on the line marked (a)) and also for
$H_{c1}(0)=2\times 10^{16}$ G (the points on line (c)). In both cases we
see that the points lie virtually on straight lines, showing that
$\epsilon\propto\Bav$. In addition, line (c) has twice the gradient of
line (a). This confirms that despite the high field strengths we are
obliged to use (see the discussion above), we find the correct scaling
of the ellipticity: $\epsilon\propto H_{c1}\Bav$. This gives us more
confidence about our results and means we can safely extrapolate to
more typical NS field strengths. We also present ellipticities for
$\zeta^3$-superconductivity (points 
along line (b)). The level of distortion in this case is very similar
to that in the $\zeta^2$ case.

Note that for both plots in figure \ref{toroidal_ellips} we consider
unstratified models, but we find that stratification makes no
discernable difference to the ellipticity results. This is the same as
for the previous plot showing ellipticities, figure
\ref{polmix_ellips}.

\subsection{Ellipticity formulae}

In figures \ref{polmix_ellips} and \ref{toroidal_ellips} we plotted
ellipticities of two-fluid stars as a function of field strength for
poloidal, toroidal and mixed fields in normal MHD, and toroidal fields
in superconductivity. In all cases we found the scaling of points on the
plots was in convincing agreement with the expected results:
$\eps\propto\Bav^2$ in the normal case and $\eps\propto H_{c1}\Bav$ in
superconductivity. In addition, for the very highest field strengths
we were able to see slight deviations from this linear result. From
the lines fitted to our data points we get quantitative relations
between ellipticity and magnetic field strength, which we present here
scaled to somewhat lower field strengths. In all cases the formulae
are for two-fluid models without stratification, but we found that
stratified stars obeyed the same relations, to a good approximation.

We begin with results for stars with superfluid neutrons and normal
protons. In the case of purely toroidal fields we have
\beq \label{norm_tor_ellip}
\epsilon=-3.0\times 10^{-6}\brac{\frac{\Bav}{10^{15}\ \mathrm{G}}}^2,
\eeq
whilst for purely poloidal fields the relation is
\beq
\epsilon=4.3\times 10^{-6}\brac{\frac{\Bav}{10^{15}\ \mathrm{G}}}^2.
\eeq
We also consider mixed fields with $3\%$ toroidal energy. This is
relatively strong for our unstratified models, but still only produces a
14\% reduction in oblateness from the purely poloidal case.

Next we consider models comprising superfluid neutrons and type-II
superconducting protons with purely toroidal magnetic fields. As
discussed in section 3.1, we have some flexibility in choosing the
function $\eta$ which governs the magnetic field. For our first choice of
function, $\zeta^2$-superconductivity, we have
\beq \label{sc_tor_ellip1}
\epsilon=-2.3\times 10^{-5}
             \brac{\frac{H_{c1}(0)}{10^{16}\ \mathrm{G}}}
             \brac{\frac{\Bav}{10^{15}\ \mathrm{G}}},
\eeq
whilst for $\zeta^3$-superconductivity the relation is
\beq \label{sc_tor_ellip2}
\epsilon=-2.6\times 10^{-5}
             \brac{\frac{H_{c1}(0)}{10^{16}\ \mathrm{G}}}
             \brac{\frac{\Bav}{10^{15}\ \mathrm{G}}}.
\eeq
\skl{We recall that these results were obtained using a density-dependent
first critical field $H_{c1}=h_c\rho_\rmp$, where $h_c$ is a constant};
in the above expressions we use the central critical field value
$H_{c1}(0)$, normalised to $10^{16}$ G. The 
equivalent volume-averaged value is
$\bar{H}_{c1}\approx 3\times 10^{15}$ G. We note that the functional
form of $\eta$ has little effect on the ellipticity relation. The
results in the superconducting case are in very good agreement with
the barotropic study of \citet{akgun_wass}; using their ellipticity
formula (71) gives a result just 5\% different from our
$\zeta^2$-superconductivity relation \eqref{sc_tor_ellip1} and 15\%
different from \eqref{sc_tor_ellip2}.

\begin{figure}
\begin{center}
\begin{minipage}[c]{0.5\linewidth}
\psfrag{Bav1}{$\displaystyle\frac{\Bav}{10^{16}\ \mathrm{G}}$}
\psfrag{ellip}{$-\epsilon$}
\psfrag{superconductivity}{superconductivity}
\psfrag{normal MHD}{normal MHD}
\includegraphics[width=\linewidth]{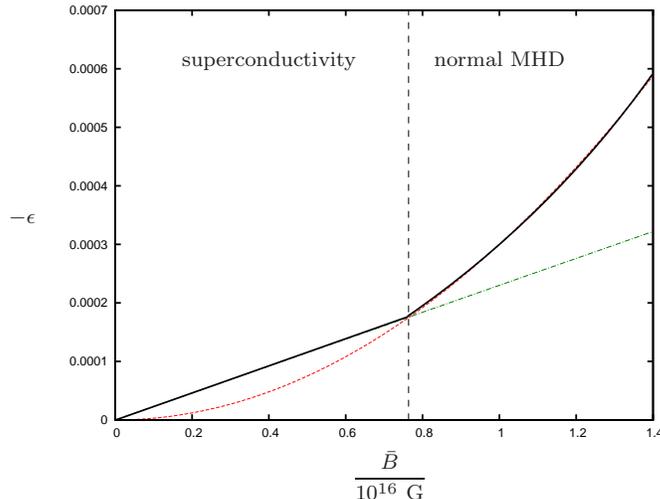}
\end{minipage}
\caption{\label{sc_to_normal}
         Below the (volume-averaged) second critical field
         $\bar{H}_{c2}$ a NS has type-II 
         superconducting protons and its ellipticity scales with
         $H_{c1}\Bav$ --- considerably higher than the normal-MHD
         curve plotted below it. Above the critical field
         superconductivity is destroyed and 
         the ellipticity has the normal-MHD scaling, quadratic in
         $\Bav$.}
\end{center}
\end{figure}

We summarise these last two extrapolated formulae in figure \ref{sc_to_normal},
where we show the expected variation of ellipticity with field
strength in a two-fluid NS model with a toroidal magnetic field. We
assume that below the volume-averaged value of the second critical
field $\bar{H}_{c2}\approx 10^{16}$ G the 
protons are superconducting, whilst above it they are normal. The
neutrons are a superfluid in both cases. In the superconducting
regime, $\epsilon\propto\Bav$ and the star is more distorted than one
would expect from normal-MHD models. As the field strength is
increased superconductivity is broken and the protons obey normal MHD, with
$\epsilon\propto\Bav^2$.

Note that superconductivity can make a huge difference to the
ellipticities predicted for most pulsars \citep{wasserman}; the
discrepancy from normal MHD scales with $\bar{H}_{c1}/\Bav$. Let us
take a pulsar 
with a volume-averaged field $\Bav=10^{13}$ G, and assume this field is purely
toroidal (since our only superconducting models are for this case). If the
protons form a normal fluid we may apply equation
\eqref{norm_tor_ellip} to predict 
that its ellipticity will be $-3.0\times 10^{-10}$; if they are a
type-II superconductor then equation \eqref{sc_tor_ellip1} predicts that the
ellipticity of the pulsar will be $-2.3\times 10^{-7}$ --- a factor of
around 800 larger.

\section{Discussion}

This paper, together with a companion paper \cite{GAL}, presents the
first results on the equilibria of multifluid neutron stars with
magnetic fields. Although it has been thought for decades that the
fluid interiors of \skl{all but the youngest} neutron stars are likely
to comprise  
(predominantly) a neutron superfluid and a magnetised proton fluid,
no previous studies have constructed equilibria of this kind. A
related problem is the role of stratification in NSs. This renders the
stellar matter non-barotropic and prevents the construction of
single-fluid equilibria using the usual Grad-Shafranov equation
\citep{reis_strat}. Virtually no studies have attempted to confront
this extra difficulty; to our knowledge the only previous studies of
magnetic equilibria in stratified stars are by Braithwaite --- although
these adopt an ideal-gas EOS more suited to main-sequence stars
\citep{braith_nord,braithtorpol} --- and the recent paper by
\citet{mastrano}. Finally, in most neutron stars the
protons are likely to form a type-II superconductor, which
significantly changes the form of the magnetic force from the familiar
Lorentz force. Once again, this effect has been neglected in most
studies; the only related work on superconducting NS equilibria is
that of \citet{akgun_wass}, although these are single-fluid models
which do not have a separate neutron superfluid.

Within this paper we explore some of these issues. In
the normal-MHD case, we show how the equations for a magnetised
barotropic star may be generalised quite straightforwardly to a
two-fluid model in which each fluid species obeys its own polytropic
relation. By choosing different proton and neutron polytropic indices,
we have a natural way of introducing stratification into the
models. \skl{For simplicity we assume that all the stellar matter is
  multifluid, whereas NSs will also have regions which behave as a
  single fluid; when purely multifluid stars are better understood, it
  will be a logical future step to account for the effect of having
  these different fluid regions.}

\skl{The equilibrium equations governing our stellar equilibria
  feature a number of apparently arbitrary magnetic functions, related
  to the strength of the magnetic field and its poloidal and toroidal
  field components. We argue that there are actually a number of
  restrictions on these functions, with little freedom in choosing
  them --- and hence limited scope for producing qualitatively
  different equilibria from those we present here. Given this, we
  believe our results are quite generic to multifluid stars.}
  
We also discuss the equations governing equilibria with type-II
superconducting protons, starting from the relations derived in detail
by \citet{GAS2011}. The main difference from the normal-MHD two-fluid
case is in the form of the magnetic force, which depends on the usual
magnetic field $\bB$ but also on the first critical field $H_{c1}$,
related to the presence of quantised fluxtubes. We use an approach inspired by
the derivation of the Grad-Shafranov equation (see, e.g.,
\citet{1f_eqm}) to try to produce equilibrium equations that may be
readily solved. As in the normal-MHD case, we arrive at a general
equation governing the magnetic field, at which point one must
specialise to different field configurations to get a solution. We
remark that one important step from the normal-MHD derivation does not
carry through to the superconducting case (and so may lead to
significant differences between the two states), but we postpone a detailed
discussion of this point to a later paper. For a purely toroidal
magnetic field however, a solution may be obtained quite easily, so we
specialise to this case in this paper.

In all cases, the equilibrium equations are solved numerically using
an iterative procedure, which is a variation of the Hachisu SCF method
\citep{hachisu}. This method is quite versatile and allows us to
consider extremely strong magnetic fields, in contrast with the study
reported in \citet{GAL}, where the magnetic field is regarded as a
perturbation on a spherical background star. The good agreement
between the results of this paper and \citet{GAL}, however, vindicate
the latter work's perturbative ansatz.

One main aim of this paper is to explore the role of
stratification in a magnetised neutron star. We increase the 
stratification of the star by increasing the value of the proton-fluid
polytropic index $N_\rmp$, whilst fixing $N_\rmn=1$. For a purely
poloidal field, this results in a larger closed-field line region,
with the `neutral line' (where the field strength vanishes) moving
inwards; see figure \ref{Bpol_vs_Np}. In mixed-field configurations
the toroidal field component is contained in this region, and is
consequently larger in stratified stars. Even in the stratified case,
however, the energy in the poloidal component is always considerably
greater than 
that in the toroidal component. This is because the parameter which
increases the toroidal-component strength also \emph{decreases} the
size of the region occupied by this component (figure
\ref{increase_a}). These results are in good agreement with those of
\citet{GAL}. \skl{Finally, for magnetic ellipticity relations our results
  suggest that stratification is not very important; there is little
  difference from the single-fluid barotropic relations.}

We have not investigated the stability of any configurations presented
here, \skl{although purely toroidal fields are unstable in both stratified
  and unstratified stars}
\citep{goos_tayler}, and it seems likely that the same will be true for
purely poloidal fields, as in the unstratified case
\citep{markey}. This leaves mixed poloidal-toroidal fields as the only
candidates for stable configurations, and hence the best models for NS
magnetic fields. Although our configurations have quite small
percentages of magnetic energy in the toroidal component, due to the
small volume of the star it occupies, the magnitude of the two
field components can be comparable. This suggests that these models
could indeed be 
stable \citep{wright}. Note that \citet{braithtorpol} suggested that a
considerably higher percentage of toroidal-field energy was needed for
stability. That work considered very different stellar models, however,
more applicable to main-sequence stars than neutron stars. \skl{There
  is no consensus, as yet, about what percentages of each field
  component are likely to exist in real NSs.}

Our other results concern equilibrium configurations of neutron stars
with superfluid neutrons and type-II superconducting protons. In this
initial study we specialise to purely toroidal magnetic fields, even
though the equilibria we generate are likely to be unstable
\citep{akgun_wass}. From our results we are able to produce the first
formulae for magnetic ellipticities of superconducting NSs based on a
two-fluid model. Our results suggest that the distortion is
around 50\% greater than the estimate used by \citet{cutler_prec}
(having accounted for the fact that we also use a larger value for the
critical field $H_{c1}$). There is very good agreement, however, with the
ellipticity formula (equation 71) from \citet{akgun_wass}. If a
substantial proportion of protons in a typical NS interior form a
type-II superconductor, then ellipticity formulae based on models
in normal MHD (e.g. \citet{haskell} and \citet{1f_eqm}) greatly
underestimate the potential magnetic distortion.

With the two-fluid model discussed in this paper, we are able to
produce a wide range of different NS equilibria, accounting for
stratification, superconductivity (with toroidal fields) and
rotation. As well as being interesting in their own right, these
equilibria could be used as background models on which to study
perturbations. This would help us understand the stability of these
various models, and their oscillation modes. The next logical step,
however, is to consider superconducting stars with purely poloidal and
mixed poloidal-toroidal magnetic fields. We hope to tackle this
problem in a future study.

\section*{Acknowledgments}

SKL acknowledges funding from the European Science Foundation (ESF)
for the activity entitled `The New Physics of Compact Stars', NA
acknowledges support from STFC in the UK through grant number
ST/H002359/1 and KG is supported by an Alexander von Humboldt fellowship
and by the German Science Foundation (DFG) via SFB/TR7. We thank the
referee Andreas Reisenegger for his detailed reading of this paper
and useful criticism.

\label{lastpage}

\end{document}